%% file: BF13_arxiv_v1.tex
\documentclass[11pt]{article}
\pdfoutput=1 

\usepackage{calc}
\usepackage{rotating}
\usepackage[english]{babel}
\usepackage{graphicx}
\usepackage{subfig}
\usepackage{float}
\usepackage{amsmath}
\usepackage{amssymb}
\usepackage{amsthm}
\usepackage{latexsym}
\usepackage{dcolumn}
\usepackage{geometry}
\usepackage{color}
\usepackage{hyperref}
\usepackage{mciteplus}

\def\gtwid{\mathrel{\raise.3ex\hbox{$>$\kern-.75em\lower1ex\hbox{$\sim
$}}}}
\def\vio{\mathrel{\hbox{$E$\kern-.60em\hbox{$/
$}}}}
\input macros_BF_13.tex

\restylefloat{figure}

\begin{document}

\title{Explaining the Fermi Galactic Centre Excess in the CMSSM}

\author{Andrew J.~Williams$^1$\\[2ex]
\small $^1$ National Centre for Nuclear Research,  Ho{\. z}a 69, 00-681 Warsaw, Poland \\
\bigskip \\
\url{Andrew.Williams@ncbj.gov.pl}}


\date{}

\maketitle

\begin{abstract}
We present an analysis of the compatibility between the Galactic Centre Excess (GCE) and the Constrained MSSM (CMSSM). 
We perform a global fit to the relevant experimental data including the GCE taking into account the systematic uncertainties.
We find that the CMSSM is able to account for the GCE and maintain agreement with the other experimental searches,
providing the first example that the GCE can be explained in the framework of universal supersymmetry.
We map out the region compatible at $2\sigma$ and comment on its phenomenology.
We find that for the CMSSM to explain the GCE the solution must lie close to the existing limits from LUX, IceCube and the LHC.
We show that this provides definite predictions for Run 2 of the LHC, Xenon-1T and future observation with IceCube.
Thus there exists the exciting possibility that the CMSSM could be observed in four distinct experimental channels over the next few years,
which would be a striking signature for universal supersymmetry.
\end{abstract}
\newpage
\section{Introduction}
The Galactic Centre (GC) represents one of the most promising targets in the search for dark matter (DM).
The GC is thought to contain the largest density of DM in the Milky Way and is thus possibly the brightest source of gamma rays from DM annihilation.
It is therefore of great interest that analyses\cite{Goodenough:2009gk,Boyarsky:2010dr,Hooper:2010mq,Hooper:2011ti,Abazajian:2012pn,Linden:2012iv,Gordon:2013vta,Hooper:2013rwa,Abazajian:2014fta,Daylan:2014rsa,Calore:2014xka,Zhou:2014lva} of the Fermi-LAT data have consistently found an excess over the background estimate in the 2-5\gev\ range.
The purported galactic centre excess (GCE) is statistically significant and compatible with a contribution from annihilating DM.
The situation is complicated by the uncertainty in the expected density profile of DM in the GC and significant systematic uncertainty in the background\cite{Calore:2014xka,Zhou:2014lva,Murgia2014web}.
It has been suggested that the excess could be due to astrophysical sources such as unresolved millisecond pulsars\cite{Abazajian:2010zy,Calore:2014oga,Gordon:2013vta} or an injection of cosmic rays\cite{YusefZadeh:2012nh}.
Recent analyses provide some evidence that the excess is made of unresolved point sources\cite{Bartels:2015aea,Lee:2015fea}.
On the other hand there is some highly tentative evidence that compatible signal may be present in Reticulum II which would bolster the claim that the signal is DM\cite{Geringer-Sameth:2015lua}.
Currently the situation remains unclear but it has been reported by Fermi collaboration that when considering several different background model the overall fit 
is improved by the addition of a spherically symmetric component with a spectrum compatible with DM annihilation\cite{Murgia2014web}.

This tantalizing signal has sparked many studies to find solutions compatible with this excess in phenomenological scenarios of new physics including both the Minimal Supersymmetric Standard Model (MSSM)\cite{Caron:2015wda,Bertone:2015tza,Freese:2015ysa}
and Next to Minimal Supersymmetric Standard Model (NMSSM)\cite{Bi:2015qva,Cao:2015loa,Cheung:2014lqa,Guo:2014gra,Cahill-Rowley:2014ora}.
It has become clear that allowing for the relatively large systematic uncertainties on the background model a range of masses and annihilation final states remain compatible with excess,
covering a mass range from $15\gev\ \lsim \mchi \lsim 300\gev$\cite{Agrawal:2014oha}.
Given this freedom it is perhaps unsurprising that the signal can be accommodated in the large parameter spaces of the MSSM and NMSSM.

In this paper we explore the possibility that the GCE finds its origin in the Constrained Minimal Supersymmetric Standard Model  (CMSSM). 
The CMSSM is still a popular framework for studying Supersymmetric (SUSY) models described by only a few unified parameters\cite{Kane:1993td}.
This presents an interesting scenario since the parameter space of the CMSSM is already highly constrained 
and observations in different channels (direct detection, LHC, indirect detection) are more highly correlated than in phenomenological models.
We perform a global frequentist analysis of the CMSSM including a fit to the GCE.
We find that, taking into account the large systematic uncertainty in the backgrounds, the CMSSM is able to explain the GCE although in some tension with other constraints.
This provides the first example that the GCE can be explained in the framework of universal SUSY.

The CMSSM has been previously heavily investigated in global fits\cite{Baer:2011ab,*Baer:2012uya,Kadastik:2011aa,Cao:2011sn,Ellis:2012aa,Bechtle:2012zk,Balazs:2013qva,
Fowlie:2012im,Akula:2012kk,Buchmueller:2012hv,Strege:2012bt,Cabrera:2012vu,Kowalska:2013hha,Dighe:2013wfa,Buchmueller:2013rsa,Roszkowski:2014wqa,Bechtle:2015nua} and has become increasingly constrained by direct searches of SUSY particles and the measurement of the Higgs boson mass at the LHC, 
as well as direct detection experiments placing limits on the DM properties.
Overall with the exception of the anomalous magnetic moment of the muon the CMSSM has remained compatible with the current observational data\footnote{In fact a recent analysis claiming that the CMSSM provides a poor fit to the data\cite{Bechtle:2015nua} hinges essentially only on the \gmtwo\ anomaly which so far remains unconfirmed.}.
Our recent Bayesian study of the CMSSM parameter space found a preference for solutions with a ~1\tev\ predominantly higgsino DM 
but that solutions achieving the correct relic abundance though $A^0$-resonance annihilation and stau-conannihilation still contribute to the posterior probability distribution\cite{Roszkowski:2014wqa}.
While the focus point region is less favored due to the limit on the spin-independent scattering cross section coming from LUX.

The paper is organized as follows. In \refsec{sec:method} we describe the scanning methodology, experimental constraints included in the likelihood function and numerical tools used.
In \refsec{sec:results} we present the results of the scan characterizing the compatible region, including prospects for future detection at other experiments.
We give a summary and conclusions in \refsec{sec:conclusion}

\section{Constraints and methodology}
\label{sec:method}
Our aim is to explore the regions of parameter space compatible with the current experimental data including the GCE.
Our previous studies followed a Bayesian approach outlined in\cite{Fowlie:2011mb,Roszkowski:2012uf,Fowlie:2012im,Kowalska:2013hha,Roszkowski:2014wqa}, 
in this work we allow for a frequentist interpretation and present results in terms of a $\chi^2$ function.
We construct the total $\chi^2$ from a global likelihood function incorporating all of the relevant experimental data.
We account for positive measurements with a Gaussian likelihood function and upper limits are implemented as described in\cite{deAustri:2006pe}.
The experimental constraints are summarized in Table.~\ref{tab:exp_constraints}.
We delay description of the $\chi^2$ contribution from the GCE to the end of this section.
The Higgs boson constraint is implemented using \higgssignals\cite{Bechtle:2013xfa} and \higgsbounds\cite{Bechtle:2008jh,Bechtle:2011sb,Bechtle:2013wla} using \feynhiggs\cite{Heinemeyer:1998np} to calculate the Higgs mass and observables.
The contribution arising from LUX\cite{Akerib:2013tjd} is applied following the procedure developed in\cite{Kowalska:2014hza} to calculate likelihood map in plane of \mchi\ and the spin-independent scattering cross section with  the proton, \sigsip.
A similar approach was taken for direct SUSY searches at the LHC and is described in detail in\cite{Roszkowski:2014wqa}.
The upper limit on the spin-dependent scattering cross section, \sigsdp, from the IceCube search for dark matter annihilating in the sun\cite{Aartsen:2012kia}, was calculated as follows.  
For each point the dominant annihilation final state was found and the upper limit for that final state and neutralino mass interpolated from\cite{Aartsen:2012kia}.
The $\chi^2$ contribution was then found using the upper limit procedure of\cite{deAustri:2006pe}.
Where a limit was not quoted for the annihilation final state of the model the $W^+W^-$ final state was used instead.
This provides only an approximation to the true $\chi^2$ since in general the neutralino does not annihilate to a single final state.
We take the same approach for the upper limit on \sigv\ from the Fermi-LAT observations of dwarf galaxies\cite{Ackermann:2015zua}. 

We take the constraint on the relic abundance of DM, \abundchi, as an upper limit only and allow scenarios where the freeze-out relic abundance is below the value given by Planck\cite{Ade:2013zuv}.
There are two frameworks in which to interpret scenarios where the calculated relic abundance is less than the measured value.
The first is that the neutralino does not account for the entire DM population as occurs in two component DM models.
In this case we apply the scaling ansatz and apply a scaling factor $\eta = \Omega h^2_{\textrm{FO}}/\Omega h^2_{\textrm{Planck}}$, where $\Omega h^2_{\textrm{FO}}$ is the calculated relic abundance from freeze-out,  everywhere that the DM density appears.
Practically this means that DM indirect detection observables are rescaled by a factor $\eta^2$ and direct detection by $\eta$.
In the second framework we do not assume a two component DM scenario and do not apply a rescaling factor to the DM observables.
Instead we allow for the situation where an under abundant neutralino has its abundance regenerated by some other mechanism such as late decays of another particle. 
Such a scenario has previously been explored in the framework of the phenomenological MSSM\cite{Williams:2012pz} and acts to strengthen the limits coming from both direct and indirect detection.
In the CMSSM we find that viable regions with \abundchi\ significantly less than 0.1199 occur only in the second framework and so only present those results here.

Since the GCE depends intimately on the DM density profile in the inner Galaxy we parameterize the halo profile using the generalized Navaro-Frenk-White (NFW) profile\cite{Navarro:1995iw}.
The generalized NFW profile is given after some manipulation by
\be
\rho(r) = \rho_0 \left(\frac{r}{r_0}\right)^{-\gamma}\left(\frac{(1 + \frac{r_0}{r_s})}{(1 + \frac{r}{r_s})}\right)^{3-\gamma},
\ee
where $r$ is the radius from the GC, $\rho_0$ is the local DM density, $r_0 = 8.5\kpc$ is the solar radius, $r_s$ is the scale radius which we set to $20\kpc$ and $\gamma$ is the slope parameter.
We allow for two parameters of the halo profile to vary, $\rho_0$ and $\gamma$.
$\rho_0$ enters into the limits from local direct detection experiments as well the limit from IceCube.
We therefore apply a rescaling to \sigsip\ and \sigsdp\ to account for this when comparing to the limits.
For the GCE we calculate the $J$-factor for a range of different $\gamma$.
During the scan we then interpolate these results and apply a scaling according to $\rho_0^2$ to get the correct $J$-factor for each sampled point.
 
\begin{table}[t]
\begin{center}
\begin{tabular}{|c|c|c|c|c|}
\hline
Constraint & Mean/Limit & Exp. Error & Th. Error & Ref. \\
\hline
Higgs sector & See text. & See text. & See text. & \cite{Bechtle:2013xfa,Bechtle:2008jh,Bechtle:2011sb,Bechtle:2013wla} \\
\hline
Direct SUSY searches & See text. & See text. & See text. & \cite{Drees:2013wra,*Barr:2003rg,*Cheng:2008hk,*Cacciari:2005hq,*Cacciari:2008gp,*Cacciari:2011ma,*deFavereau:2013fsa,*Lester:1999tx,*Read:2002hq} \\
\hline
\sigsip\ & See text. & See text. & See text. & \cite{Akerib:2013tjd}\\
\hline
\sigsdp\ & See text. & See text. & See text. & \cite{Aartsen:2012kia}\\
\hline
\sigv\ & See text. & See text. & See text. & \cite{Ackermann:2015zua}\\
\hline
\abunchi\ & $< 0.1199$ & 0.0027 & 10\% & \cite{Ade:2013zuv}\\
\hline
\sinsqeff\ & 0.23155 & 0.00015 & 0.00015 &  \cite{Agashe:2014kda} \\
\hline
$\brbxsgamma\times 10^4$ & 3.43 & 0.22 & 0.21 & \cite{bsgamma} \\
\hline
$\brbutaunu \times 10^4$ & 0.72 & 0.27 & 0.38 & \cite{Adachi:2012mm} \\
\hline
\delmbs\ & 17.761~ps$^{-1}$ & 0.022~ps$^{-1}$ & 2.400~ps$^{-1}$ & \cite{Agashe:2014kda} \\
\hline
$M_W$ & $80.385\gev$ & $0.015\gev$ & $0.015\gev$ & \cite{Agashe:2014kda} \\
\hline
$\brbsmumu \times 10^9$ & 2.9 & $0.7$ & 10\% & \cite{Aaij:2013aka,Chatrchyan:2013bka} \\
\hline
\end{tabular}
\caption{The experimental constraints used in this study.}
\label{tab:exp_constraints}
\end{center}
\end{table}%

The scan is orchestrated by the BayesFITS package\cite{Fowlie:2012im,Kowalska:2012gs,Kowalska:2013hha,Fowlie:2013oua,Roszkowski:2014wqa} that interfaces many public tools to calculate the physical observables. 
The sampling is performed by MultiNest\cite{Feroz:2008xx} using 10000 live points.
The evidence tolerance is set to 0.0001 and the sampling was allowed to continue until an adequate number of points with good $\chi^2$ were collected.
$\tt SoftSusy \,v.3.3.9$\cite{Allanach:2001kg} is used to calculate the mass spectrum, while
 $\tt FeynHiggs\ v.2.10.0$ is used to calculate the higher order corrections to the Higss mass,
 along with $M_W$, \sinsqeff\ and \delmbs.
 $\tt SuperISO\ v.3.3$\cite{Mahmoudi:2008tp} is used to calculate \brbxsgamma, \brbsmumu.
 We use $\tt micrOMEGAs$\cite{Belanger:2013oya} to calculate the DM observables \abundchi, \sigsip, \sigv\ and \sigsdp.

\begin{table}[t]
\begin{center}
\begin{tabular}{|c|c|c|c|}
\hline
\footnotesize Parameter & \footnotesize Description & \footnotesize Range & \footnotesize Distribution \\
\hline
\mzero\ & \footnotesize Universal scalar mass & $0.1,10$ & \footnotesize Log \\
\hline
\mhalf & \footnotesize Universal gaugino mass & $0.1,5$ & \footnotesize Log \\
\hline
\azero\ & \footnotesize Universal trilinear coupling & $-10,10$ & \footnotesize Linear \\
\hline
\tanb\ & \footnotesize Ratio of the Higgs vevs & $3,62$ & \footnotesize Linear \\
\hline
\signmu\ & \footnotesize Sign of the Higgs/higgsino mass parameter & $+1$ or $-1$ &  \\
\hline
\hline
\footnotesize Nuisance parameter & \footnotesize Description & \footnotesize Central value & \footnotesize Distribution \\
\hline
$M_t$ & \footnotesize Top quark pole mass & \footnotesize $173.34 \pm 0.76\gev$\cite{ATLAS:2014wva} & \footnotesize Gaussian \\
\hline
$\Sigma_{\pi N}$ & \footnotesize Nucleon sigma term &\footnotesize $ 34 \pm 2\mev$\cite{Belanger:2013oya} & \footnotesize Gaussian\\ 
\hline
$\sigma_s$ & \footnotesize Strange sigma commutator &\footnotesize $42 \pm 5\mev$\cite{Belanger:2013oya} & \footnotesize Gaussian\\
\hline
$\rho_0$ &  \footnotesize Local DM density & \footnotesize $0.4 \pm 0.2$ \gev/cm$^3$\cite{Catena:2009mf} & \footnotesize Gaussian\\
\hline
$\gamma$ &  \footnotesize Slope parameter of NFW profile & $1.28 \pm 0.7$\cite{Calore:2014xka} & \footnotesize Gaussian\\
\hline
\end{tabular}
\caption{Prior distributions of the CMSSM and nuisance parameters used in the scans. All dimensionful parameters are given in\tev\ unless indicated otherwise.}
\label{tab:cmssmprior}
\end{center}
\end{table}%

Since MutliNest is primarily a Bayesian tool the prior distributions of the model and nuisance parameters must be specified.
These are summarized in Table.~\ref{tab:cmssmprior}.
We combine two samples for positive and negative $\mu$.
For a Bayesian analysis it is sufficient to include our knowledge of the nuisance parameters in their prior distributions,
in a frequentist interpretation we must include their contribution to the total $\chi^2$.
In addition to the nuisance parameters considered in our previous works, the top quark pole mass, $M_t$, and nuclear form factors, $\sigma_s$ and $\Sigma_{\pi N}$ we also allow two parameters related to the DM halo profile to vary.
These are the local DM density $\rho_0$ and the NFW slope parameter $\gamma$ described above.
We do not include $\rho_0$ or $\gamma$ in the likelihood function and instead allow them to vary freely according to a gaussian prior.
For $\rho_0$ estimates range between around $0.2-0.5\gev/\textrm{cm}^3$\cite{Bovy:2012tw,Read:2014qva} we choose a central value of 0.4\gev/cm$^3$\cite{Catena:2009mf} but allow an uncertainty of 0.2\gev/cm$^3$ to account for the disagreement between determinations. 
For the parameter $\gamma$ we take $\gamma = 1.28\pm0.7$ as found in\cite{Calore:2014xka}. 
We explicitly check that in the converged chains the values do not deviate from the allowed values.

Finally we account for the GCE using the residual signal after background subtraction derived in\cite{Calore:2014xka} 
which has been made publicly available in the form of a covariance matrix $\Sigma$ accounting for the correlated systematic uncertainties in the background model.
We account for the theoretical uncertainty in the DM annihilation spectrum by an uncorrelated additional 10\% uncertainty as was adopted in a recent study of the MSSM\cite{Caron:2015wda}.
The $\chi^2_{\textrm{GCE}}$ contribution is then given by
\be
\chi^2_{\textrm{GCE}} = \sum_{ij}\left(\frac{d\mu}{dE_i} - \frac{dN}{dE_i}\right)\Sigma^{-1}_{ij}\left(\frac{d\mu}{dE_j} - \frac{dN}{dE_j}\right)
\ee
where $d\mu/dE_i$ is the expected GCE flux from DM annihilations in the $i^{\textrm{th}}$ energy bin, 
$dN/dE_i$ is the measured residual flux the $i^{\textrm{th}}$ energy bin and $\Sigma^{-1}$ is the inverse of the covariance matrix.
We remark that it has been suggested recently\cite{Agrawal:2014oha,Freese:2015ysa} and in preliminary analyses by the Fermi collaboration\cite{Murgia2014web} that the systematic uncertainty in the background may be significantly larger than that considered here,
thus allowing a greater range of possible spectra that are compatible with the data.
\section{Results}
\label{sec:results}
\subsection{CMSSM}
\begin{figure}[t]
\centering
\subfloat[]{%
\label{fig:a}%
\includegraphics[width=0.46\textwidth]{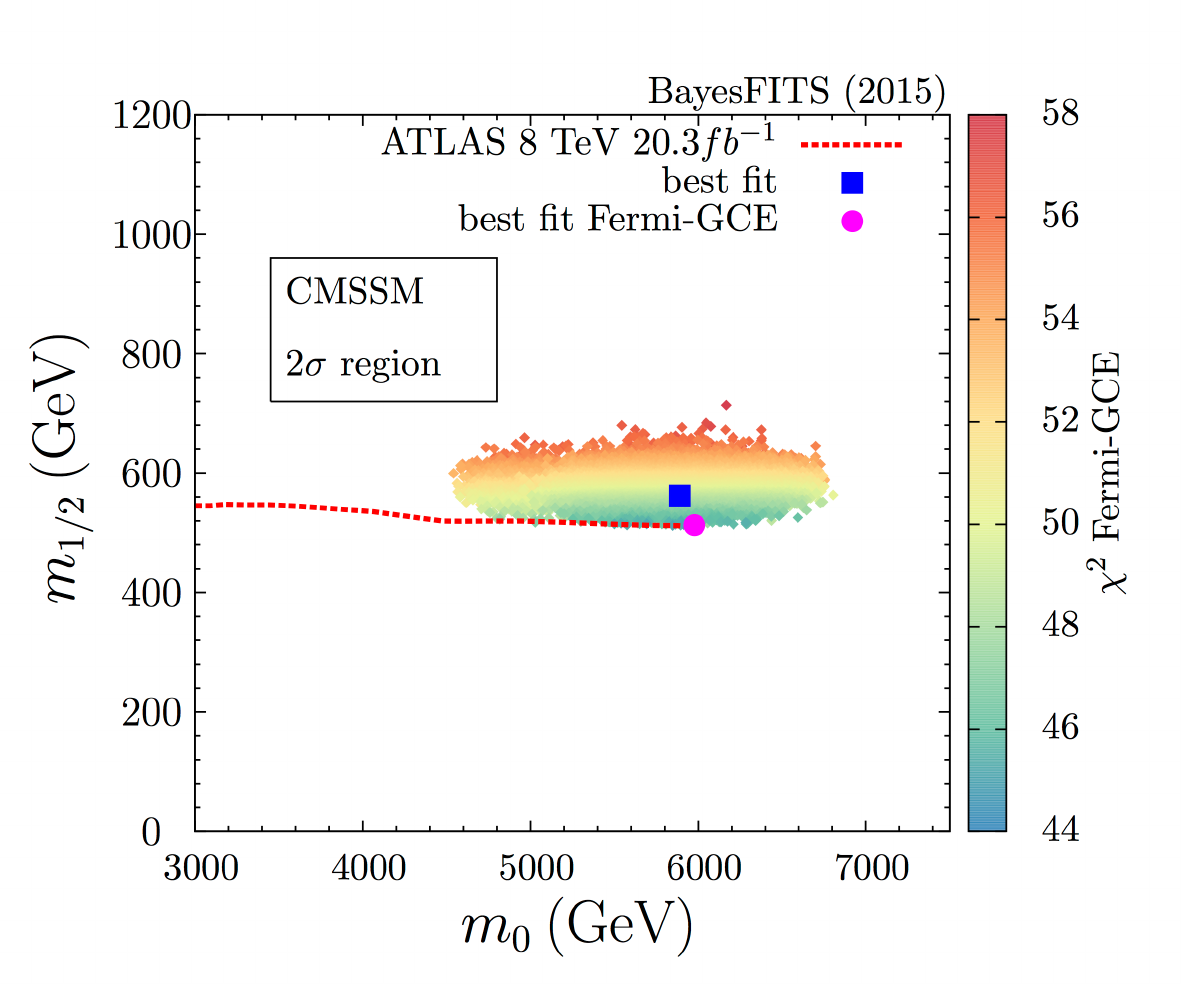}
}%
\hspace{0.007\textwidth}
\subfloat[]{%
\label{fig:b}%
\includegraphics[width=0.46\textwidth]{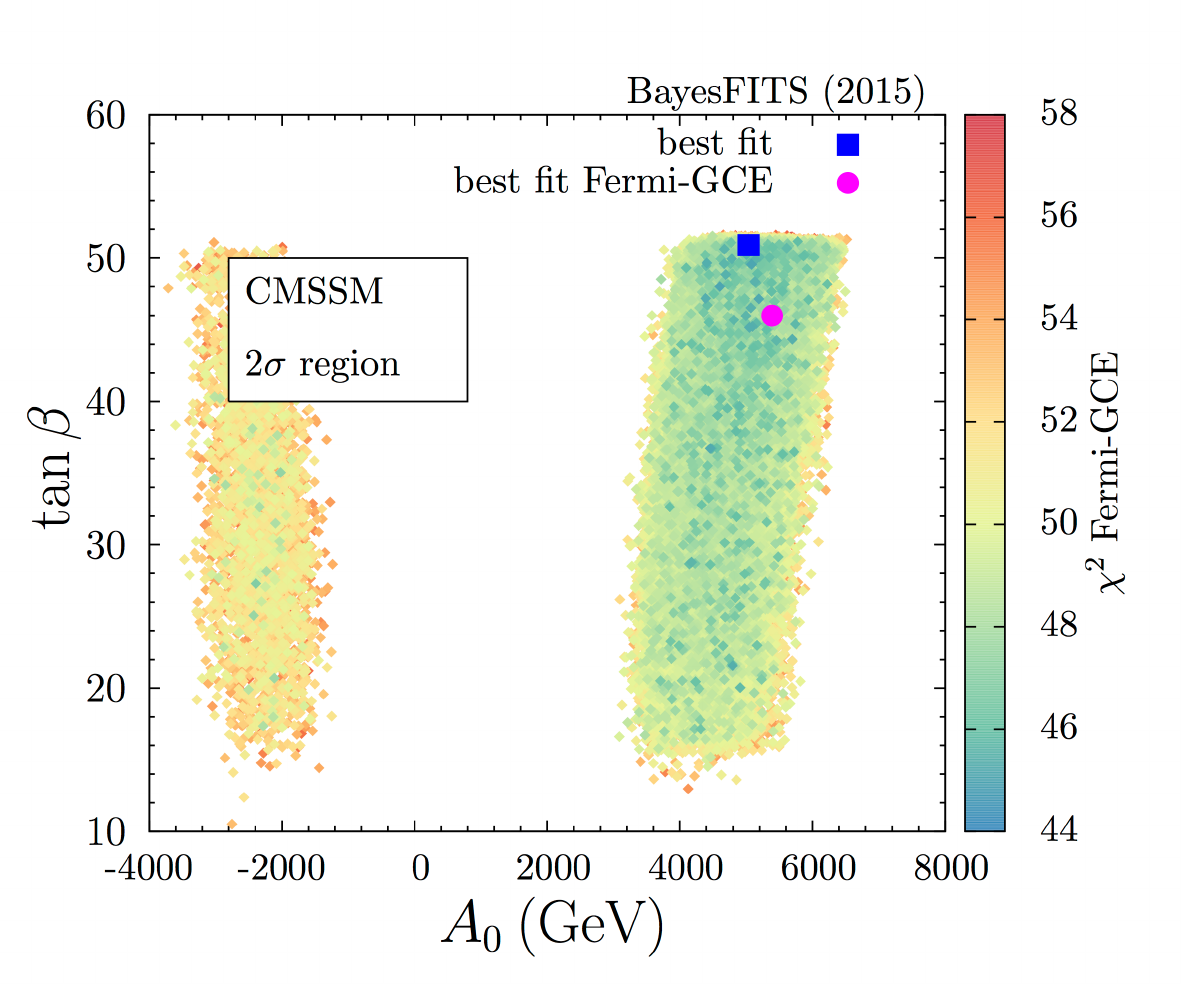}
}%
\caption{(a) The distribution of points with $\Delta\chi^2<5.99$ in the (\mzero, \mhalf) plane. The color coding displays the $\chi^2$ contribution from the fit to the GCE. 
The overall best fit point is shown as a blue square. The point with the best fit to the GCE is shown as a magenta circle. 
The ATLAS limit is shown as a dashed red line. (b) The same as (a) in the (\azero, \tanb) plane.}
\label{fig:m0m12}
\end{figure}

In this section we present the results of the scan of the CMSSM parameter space. 
Only one region is found that provides a good fit to both the GCE and other experimental limits. 
The solution appears in the focus point region, where the lightest neutralino is a mixture of the bino and higgsino.
We show in \reffig{fig:m0m12}(a) the points with $\Delta \chi^2 < 5.99$ in the (\mzero, \mhalf) plane color coded with the contribution from the fit to the GCE, $\chi^2_{\textrm{GCE}}$.
The allowed region requires a low \mhalf\ such that the neutralino is light enough to fit the GCE with the best fit to the GCE (shown as a purple circle) occurring adjacent to the current 95\% C.L. limit  from the ATLAS shown as a red dashed line.
The point with the best total $\chi^2$ is found with a somewhat larger \mhalf\ and slightly worse fit to the GCE. 
In general increasing \mhalf\ increases the neutralino mass such that the region favored by the GCE is constrained to $\mhalf < 750\gev$.
As we will discuss later this leads to a definite prediction for the current LHC Run 2.
In contrast \mzero\ is less constrained and takes values from 4500\gev\ to 7000\gev.
In \reffig{fig:m0m12}(b) we show the same points now projected in the (\azero, \tanb) plane.
Here we see a preference for positive \azero\ and a slight preference for larger values of \tanb\ with both the overall best fit and the best fit to the GCE having $\tanb > 45$.

The neutralino has mass $200\gev \lsim \mchi \lsim 280\gev$ and annihilation cross section $1.5\times10^{-26}\textrm{cm}^3/\textrm{s} \lsim \sigv \lsim 5\times10^{-26}\textrm{cm}^3/\textrm{s} $
The neutralino annihilates predominately to $t \bar{t}$ with an annihilation fraction between 0.6 to 0.85.
The remaining annihilations are to the $b \bar{b}$ final state.
This gives a fit to the GCE with a $\chi^2$ of between $44 \lsim \chi^2_{\textrm{GCE}} \lsim 58$, assuming the background model and systematics of\cite{Calore:2014xka}, 
which is comparable to that found for regions in the MSSM\cite{Caron:2015wda}.
It should be noted however that this combination of mass, cross-section and final state is compatible with other analyses\cite{Agrawal:2014oha,Freese:2015ysa} including the preliminary fits from the Fermi collaboration\cite{Murgia2014web}.

\begin{figure}[t]
\centering
\subfloat[]{%
\label{fig:a}%
\includegraphics[width=0.46\textwidth]{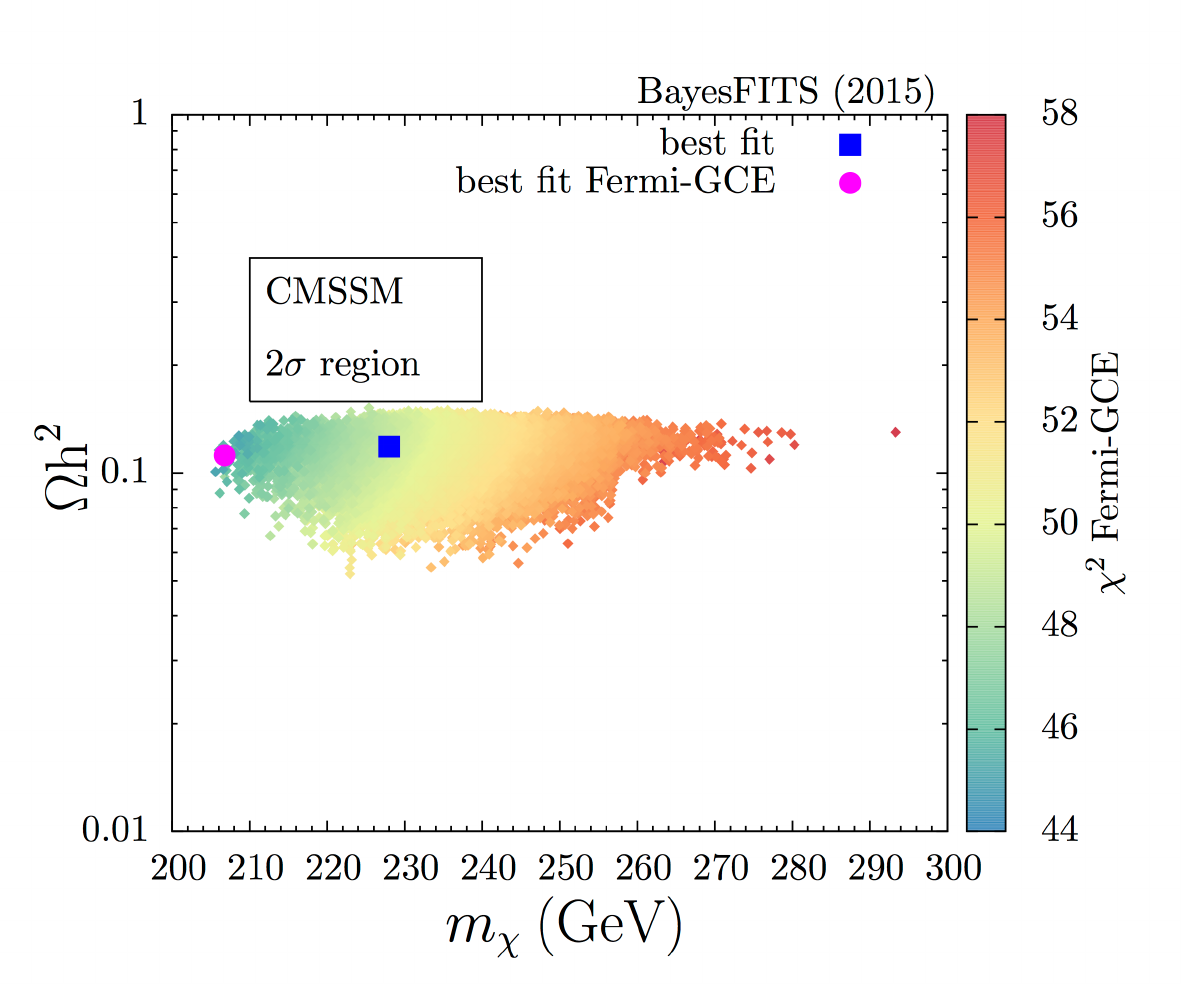}
}%
\hspace{0.007\textwidth}
\subfloat[]{%
\label{fig:b}%
\includegraphics[width=0.46\textwidth]{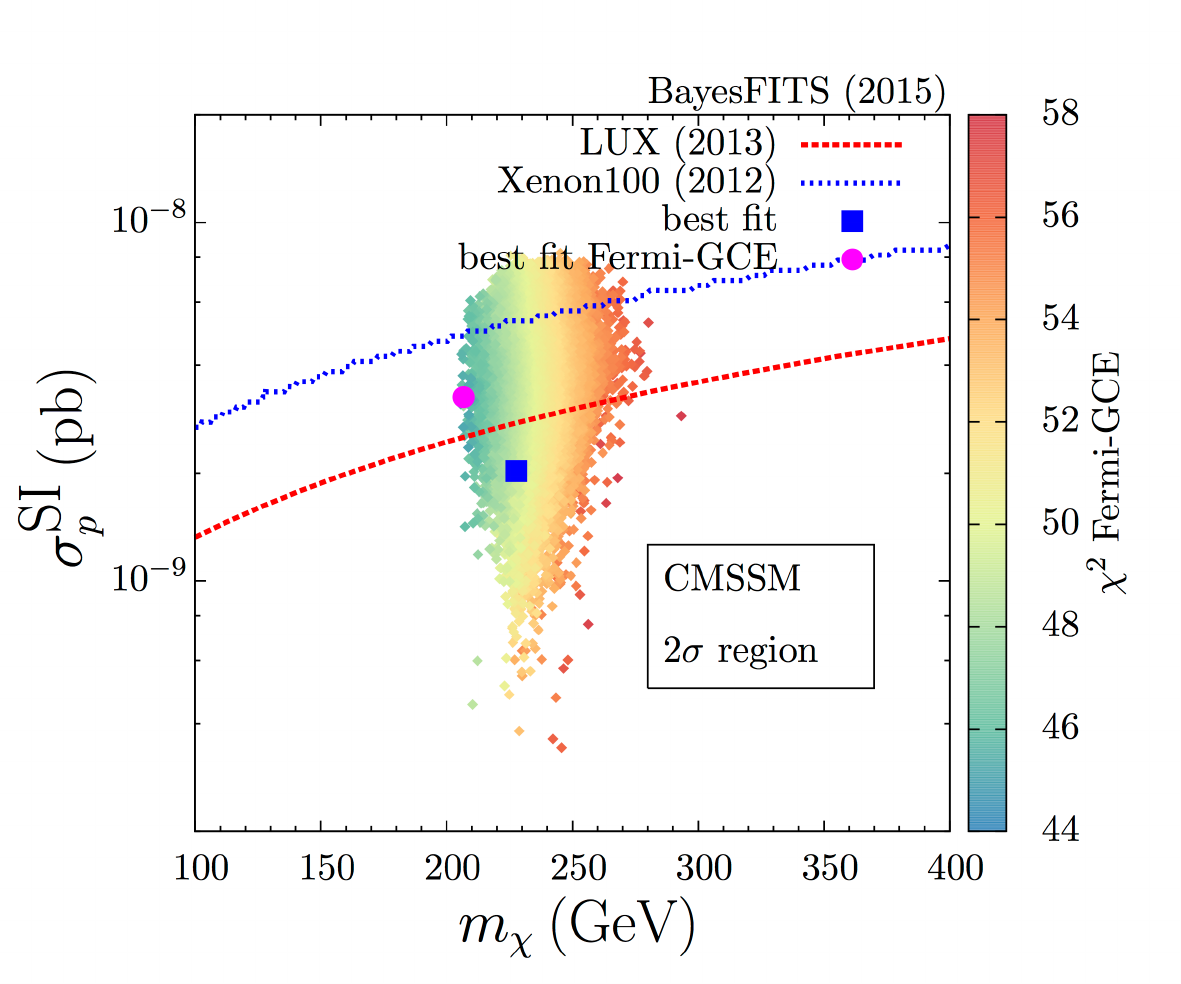}
}%
\hspace{0.007\textwidth}
\subfloat[]{%
\label{fig:c}%
\includegraphics[width=0.46\textwidth]{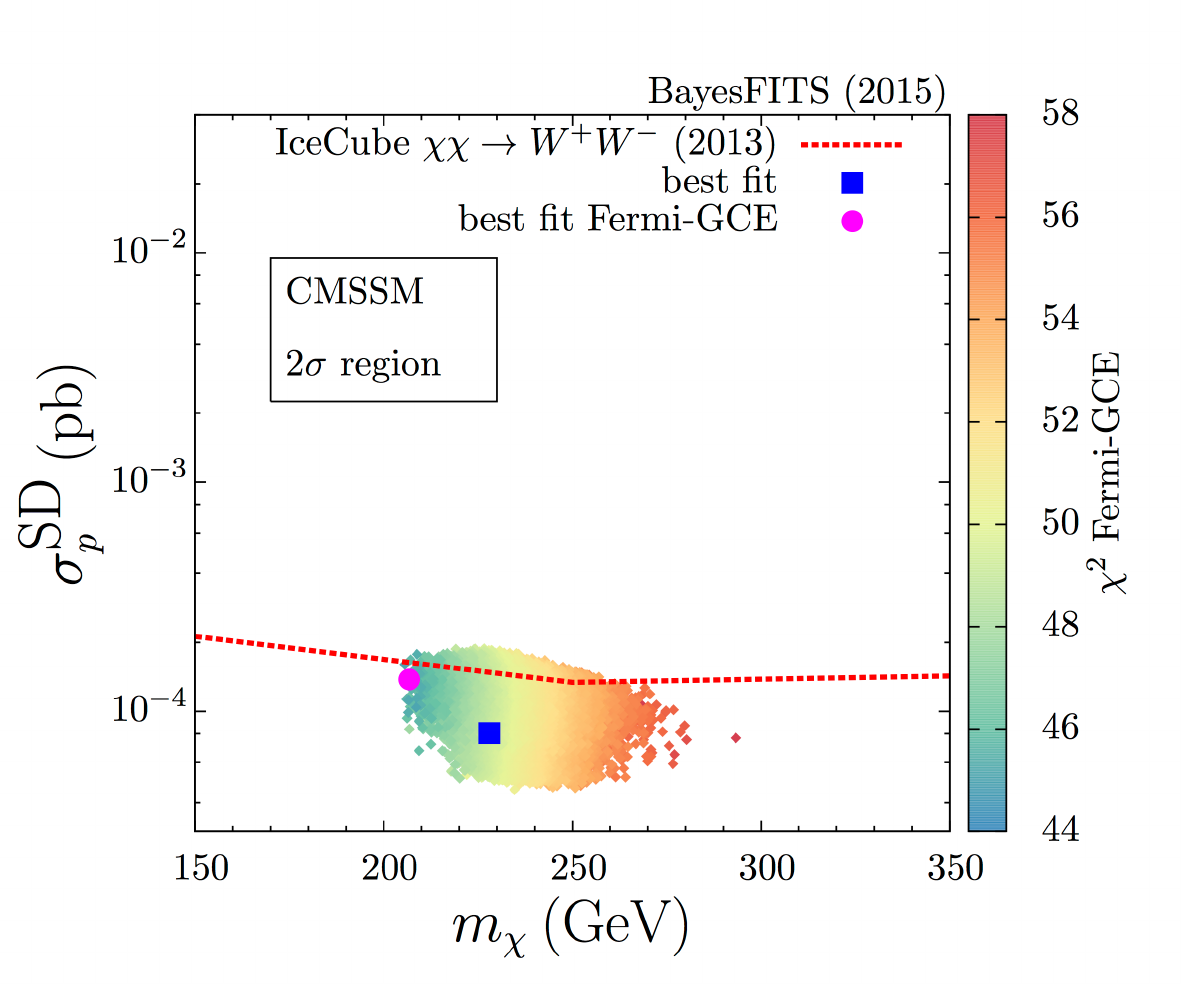}
}%
\hspace{0.007\textwidth}
\subfloat[]{%
\label{fig:d}%
\includegraphics[width=0.46\textwidth]{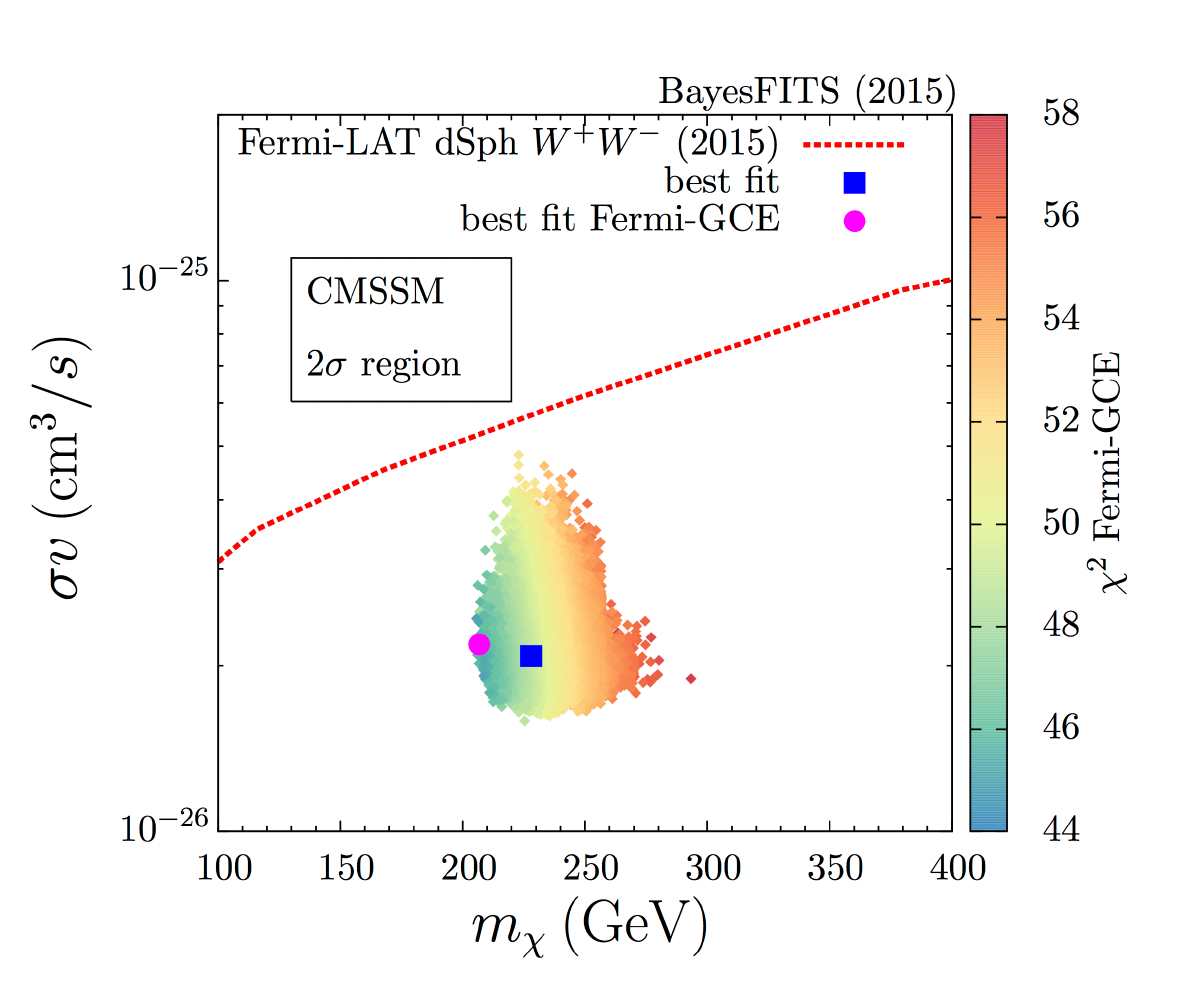}
}%
\caption{(a) The distribution of points with $\Delta\chi^2<5.99$ in the (\mchi, \abundchi) plane. Color coding is the same as \protect\reffig{fig:m0m12}.
(b) same as (a) in the (\mchi, \sigsip) plane. The 95\% C.L. limit from LUX\cite{Akerib:2013tjd,Savage:2015xta} is shown as a red dashed line.
The 90\% C.L. limit from XENON100\cite{2012arXiv1206.6288A} is shown as a blue dotted line.
(c) same as (a) in the (\mchi, \sigsdp) plane. The 90\% C.L. limit from IceCube\cite{Aartsen:2012kia} for the $W^+W^-$ final state is shown as a red dashed line.
(d) same as (a) in the (\mchi\, \sigv) plane. The 95\% C.L. limit from Fermi-LAT combined analysis of dwarf spheroidal galaxies\cite{Ackermann:2015zua} for the $W^+W^-$ final state is shown as a red dashed line.}
\label{fig:mx_omegah}
\end{figure}

In \reffig{fig:mx_omegah}(a) we show the favored region in the (\mchi, \abundchi) plane. 
Here we see that the majority of points lie close to the measured Planck value of \abundchi\ taking into account the theoretical uncertainty.
Both the overall best fit point and best fit to the GCE are compatible with the neutralino explaining the entire DM abundance without invoking any additional mechanism.
This is rather remarkable given that the Planck data was included as an upper bound only in the likelihood function. 
It is worth remarking that, since many of the points lie within the combined uncertainty of the theory calculation and experimental value, these solutions would remain valid
even if we insisted from the start that the neutralino saturate the DM abundance through thermal freeze-out.

Before discussing the other properties of the allowed region and prospects for confirmation, we briefly discuss the overall $\chi^2$ and relevance for the $2\sigma$ region.
The contributions to the $\chi^2$ for the best fit point are sumarized in Table.~\ref{tab:chisq}.
For easier comparison we have exchanged the $\chi^2$ contribution from \higgssignals\ for the simple contribution from only the Higgs mass measurement.
Note that the best fit $\chi^2=13.2$ is slightly worse than those obtained in fits without the GCE.
This is mainly a result of tension between the GCE and the limits from LUX and Run 1 of the LHC.
We again stress that the scan is driven by a global $\chi^2$ that includes the GCE data and may underestimate the systematic uncertainty in the background.
Points with a relatively poor $\chi^2_{\textrm{GCE}}$ could still be compatible with GCE.
However as a result $\chi^2_{\textrm{GCE}}$ tends to push the scan to regions in conflict with LUX and thus the $2\sigma$ region extends beyond the limit from LUX.

\begin{table}[t]
\begin{center}
\begin{tabular}{|c|c|c|}
\hline
Observable & Best fit point & $\chi^2$ contribution \\
\hline
Higgs mass &125.25 \gev\ &  0.022 \\
\hline
LHC & $\mzero=5940$ \gev, $\mhalf=562$ \gev\ & 4.2  \\
\hline
\sigsip\ & $3.4\times10^{-9}\pb$ & 3.8  \\
\hline
\sigv\ & $2.3\times 10^{-26}$ cm$^3$/s & 0.7 \\ 
\hline
\sigsdp\ & $1.1\times 10^{-4}\pb$ & 0\\
\hline
\abundchi\ & 0.108 & 1.0$^*$ \\
\hline
\brbsmumu\ &$ 3.2\times 10^{-9}$ & 0 \\
\hline
\brbsgamma\ & $3.37\times10^{-4}$ & 0.04 \\
\hline
\brbutaunu\ & $7.7\times10^{-3}$ & 0.02 \\
\hline
\mw & 80.363\gev &  1.07 \\
\hline
\sineff & 0.23151 & 0.07\\
\hline
\delmbs & 21.19 ps$^{-1}$ & 2.0 \\
\hline
\mt & 173.47 \gev\ & 0.03 \\
\hline
$\Sigma_{\pi N}$ & 33.8 \mev\ & 0.01 \\
\hline
$\sigma_{s}$ & 43.6 \mev\ & 0.1 \\
\hline
Total & - & 13.2 \\
\hline
\hline
Fermi-GCE & - & 50 \\
\hline 
\end{tabular}
\\
\footnotesize $(\ast)$ Contribution from \abundchi\ for a gaussian likelihood centred at the Planck value.
\caption{$\chi^2$ contributions from the different observables for the best fit point.}
\label{tab:chisq}
\end{center}
\end{table}%

In \reffig{fig:mx_omegah}(b) we show the allowed points in the (\mchi, \sigsip) plane. 
We show the $95\%$ \cl\ upper limit from LUX calculated using $\tt LUXCalc$\cite{Savage:2015xta} as a red dashed line.
As mentioned above the region extends above the $95\%$ \cl\ upper limit from LUX demonstrating the existing tension between LUX and the GCE.
However it is also clear that points with good fit to the GCE can also satisfy the LUX bound although they must lie relatively close to it.
The tension with LUX can be reduced by lowering the local DM density. However, this must in general come with an increase to the slope of the NFW profile such that a high enough
annihilation rate is maintained in the GC.
This therefore can have only a limited effect before either $\rho_0$ or $\gamma$ comes into conflict with their allowed ranges.
The limit from LUX could be evaded by underproducing the neutralino so that the scattering rate is scaled according to its relic abundance. 
However since the spin-independent scattering rate scales like $\eta$ and the annihilation rate in the GC scales like $\eta^2$ the fit to the GCE degrades more rapidly.
Clearly future updates from LUX and the upcoming Xenon-1T detector will have sensitivity to this region with the possibility to detect or exclude this region in the near future.
For Xenon-1T the expected exclusion sensitivity is $\sim 5\times10^{-11} \pb$\cite{2012arXiv1206.6288A} which could decisively rule out the allowed region in the absence of a signal.
On the other hand this scenario would predict of the order of hundreds of events after two years of running at Xenon-1T leading to a clear discovery.

Turning to the spin-dependent scattering rate shown in \reffig{fig:mx_omegah}(c) we see that the allowed region lies close to the current limit from IceCube.
The limit from IceCube assumes annihilation to  $W^+W^-$ final state. However, since each top quark decay also produces a $W$ we expect the flux of neutrinos to be similar\cite{Cirelli:2010xx}.
As was the case for \sigsip, the future sensitivity of IceCube will entirely probe this scenario with possibility to observe of the order of tens to a hundred of events within 5 years of operation with the full 86-string configuration.

We show in \reffig{fig:mx_omegah}(d) the annihilation cross section versus \mchi\ along with the current limit from Fermi-LAT observations of dwarf spheroidal galaxies\cite{Ackermann:2015zua} for the $W^+W^-$ final state. 
The entire favored region lies below the current limit assuming a $W^+W^-$ final state, although again this final state is not the dominant annihilation state for the DM.
Broadly speaking, however, this suggests that this scenario is consistent with the non observation of DM annihilation in dwarf galaxies by Fermi-LAT.
Again the entire region is within a factor of a few of the limit as is expected to explain the GCE.
Note that the picture is reversed with respect to the \sigsip\ case and points with lower \sigsip\ correspond to larger \sigv\ and vice-versa.
This makes the two channels highly complementary in this region such that it is not possible to hide from both experiments for long in the future.

\begin{figure}[t]
\centering
\subfloat[]{%
\label{fig:a}%
\includegraphics[width=0.46\textwidth]{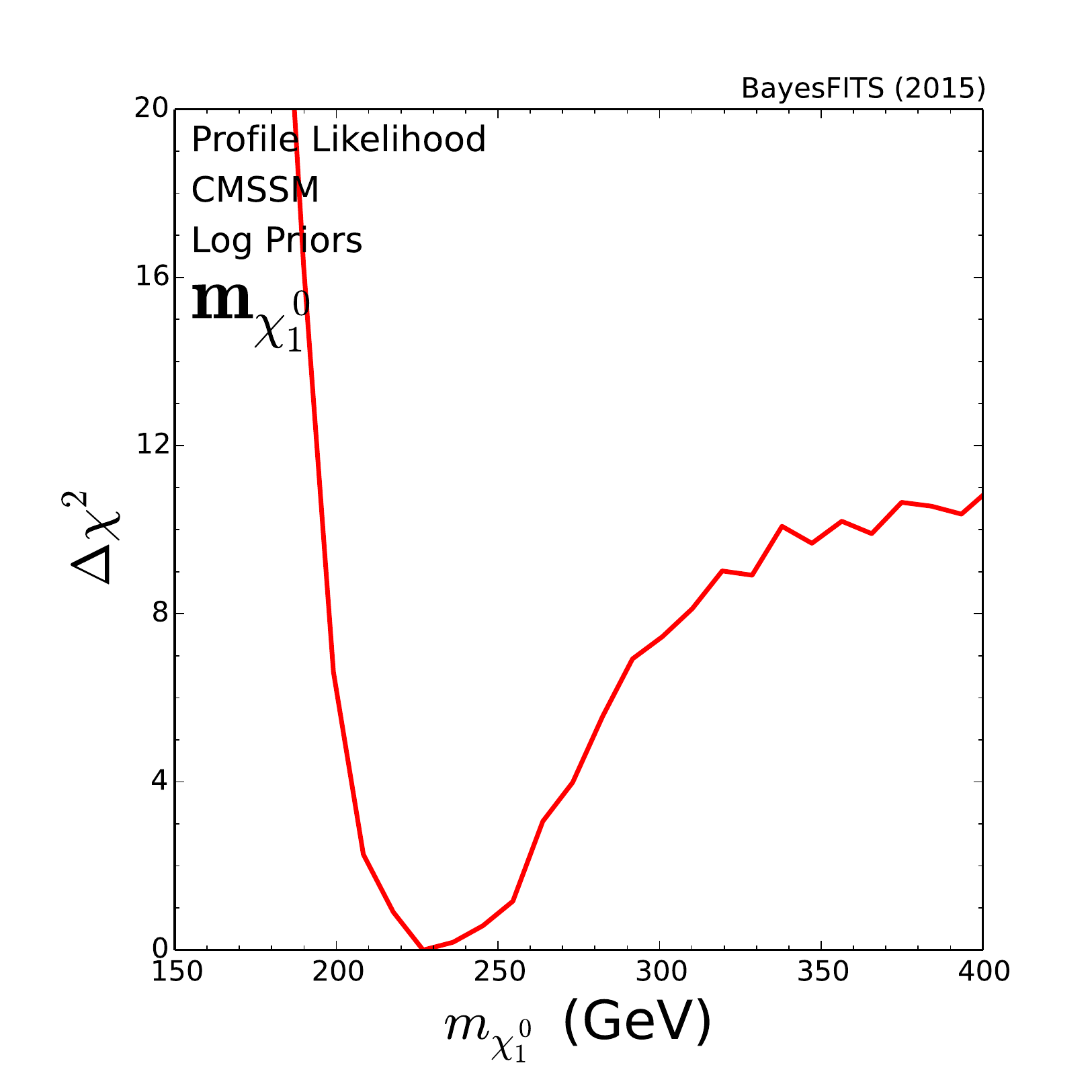}
}%
\hspace{0.007\textwidth}
\subfloat[]{%
\label{fig:b}%
\includegraphics[width=0.46\textwidth]{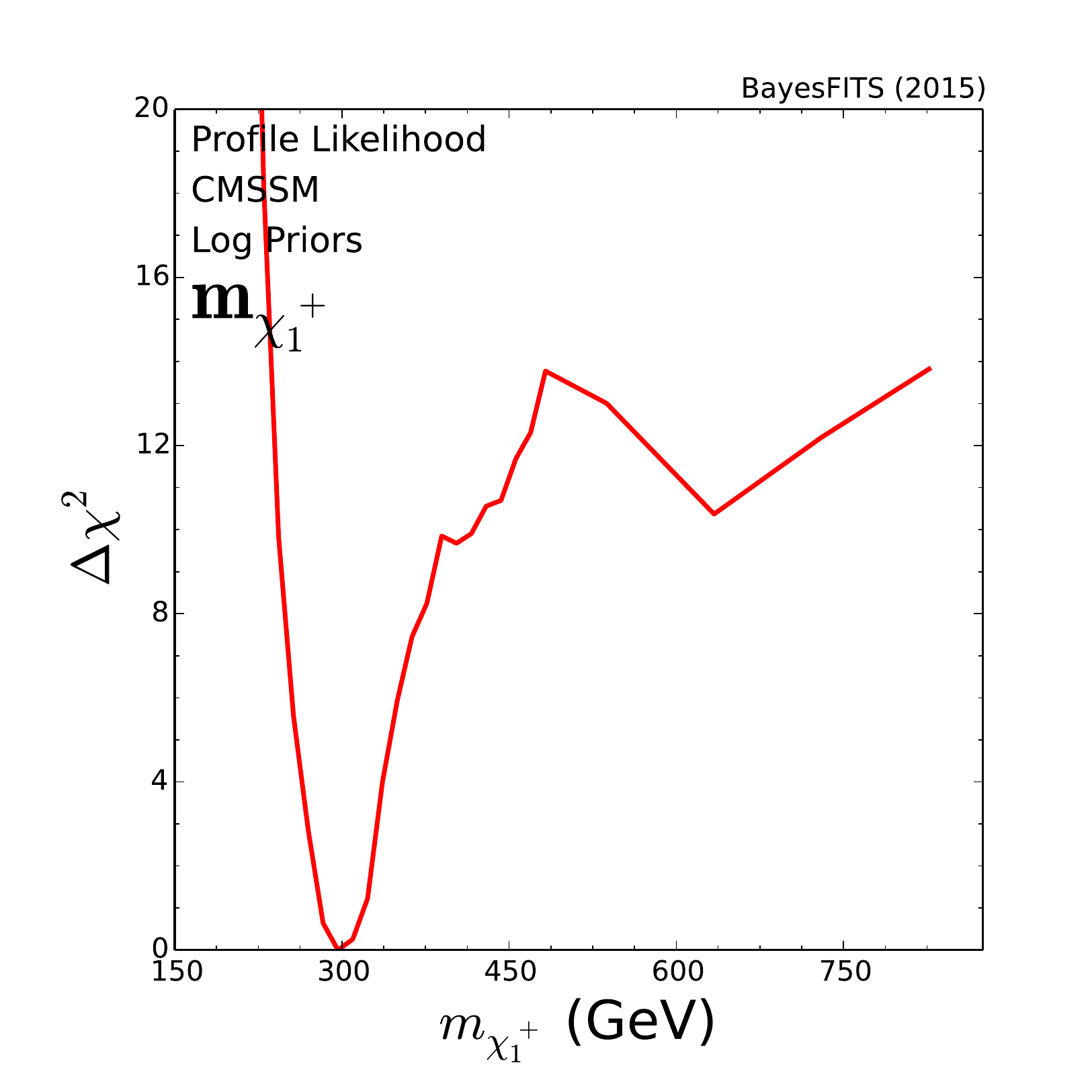}
}%
\hspace{0.007\textwidth}
\subfloat[]{%
\label{fig:c}%
\includegraphics[width=0.46\textwidth]{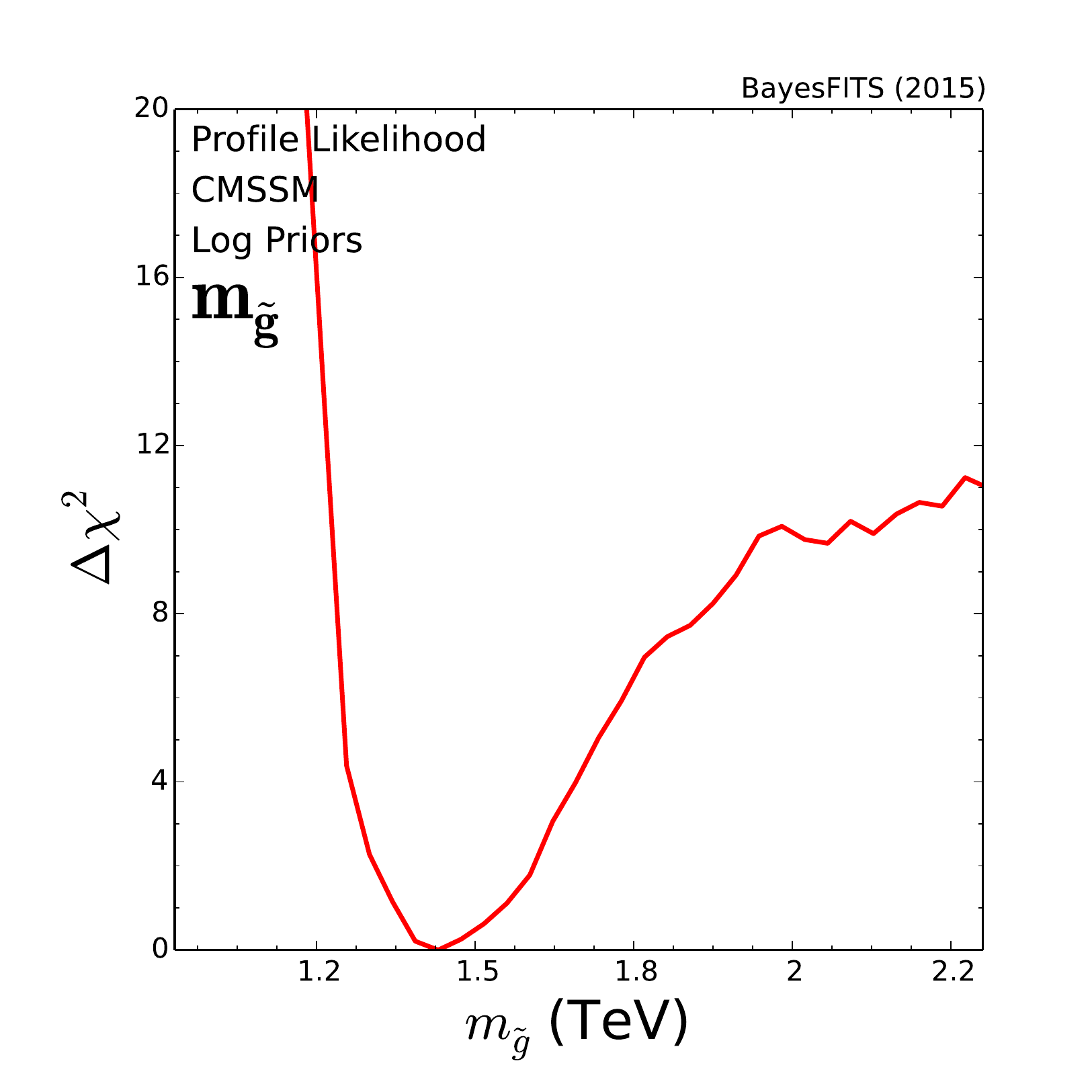}
}%
\hspace{0.007\textwidth}
\subfloat[]{%
\label{fig:d}%
\includegraphics[width=0.46\textwidth]{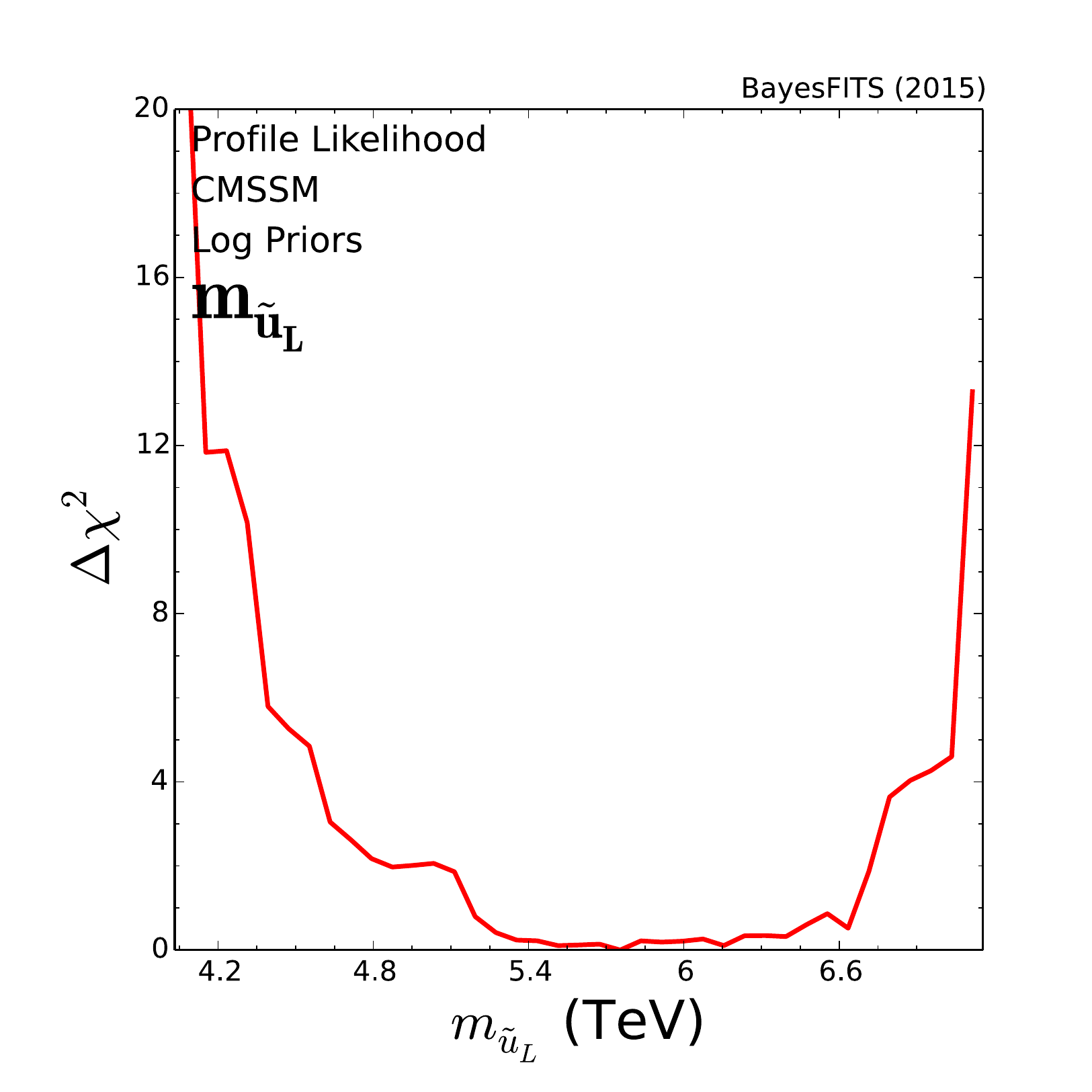}
}%
\caption{Profile likelihood as a function of (a) \mchi, (b) $m_{\chi^+_1} $, (c) \mglu\ and (d) $m_{\tilde{u}_L}$.}
\label{fig:cmssm_masses}
\end{figure}

In \reffig{fig:cmssm_masses} we show the profile likelihood distributions of the sparticle masses.
Firstly we have a relatively precise prediction for the neutralino mass shown in \reffig{fig:cmssm_masses}(a) which is directly driven by the shape of the GCE.
Connected to this we have the lightest chargino mass as shown in \reffig{fig:cmssm_masses}(b) which is mostly higgsino so enters into fit to the GCE by being related to the mixing between the bino and higgsino components of the lightest neutralino.
Due to guagino mass universality at the GUT scale the gluino mass shown in \reffig{fig:cmssm_masses}(c) is also fairly constrained, since $M_3:M_1 \approx 6:1$.
This becomes the main prediction for the LHC to arise in this scenario namely that the gluino is light and will appear in Run 2.
Extending beyond the CMSSM one could abandon gaugino mass universality at the GUT scale as was recently proposed\cite{Akula:2013ioa,Nath:2015dza}, and shown to improve agreement with the anomalous muon magnetic moment in CMSSM like models\cite{Kowalska:2015zja}, by decoupling the gluino mass parameter $M_3$ from \mhalf.
In this case the gluino could be considerably heavier and thus not appear at the LHC.
Even if one considers such a scenario the chargino can not be made heavier without effecting the properties of the DM and so this scenario could instead be searched for in electroweakino searches at the LHC.
For the squark masses since \mzero\ is large (as is required in the focus point region to achieve small $\mu$), 
Run 2 of the LHC will not be sensitive to squark production as can be seen in \reffig{fig:cmssm_masses}(d).

\section{Conclusions}
\label{sec:conclusion}
The main conclusion is that it is possible to accommodate the GCE both with and without $\abundchi \approx 0.1199$,
even in the simplest and most constrained unified SUSY model, namely the CMSSM.
There is however some tension with LUX and LHC.
If the CMSSM is the explanation for the GCE however then there is the exciting possibility that we will see signals in spin-independent direct detection experiments, neutrinos at IceCube and gluino and electroweakino production at the LHC all within the near future.

The favored region is in tension with the current limits from both LUX and the LHC, 
However an analysis of total $\chi^2$ from the conventional data, that is excluding $\chi^2_{\textrm{GCE}}$, 
showed that this region can not currently be excluded. 
The interpretation of $\chi^2_{\textrm{GCE}}$ is complicated by systematic uncertainty in the background model. 

We saw that extensions to the CMSSM, namely abandoning gaugino mass universality would lift some of the tension with the limits from Run 1 on the gluino mass.
However, even with a heavy gluino Run 2 should still probe this scenario in electroweakino searches.

Insisting on a reasonably good fit to the GCE restricts the CMSSM to a rather small region of parameter space.
This leads to a number of definite predictions if the CMSSM is to explain the GCE.
These are
\begin{itemize}
\item The mass of the lightest neuralino, \mchi, should be between 200\gev\ and 280\gev, while for the lightest chargino $250\gev \lsim m_{\chi^+_1} \lsim 350\gev$. 
The gluino mass should lie between $1.3\tev \lsim \mglu \lsim 1.7\tev$. These are well within the discovery potential for the LHC Run 2.
\item The squarks should be heavy $\gsim 4.5\tev$ and thus outside of the reach of LHC Run 2.
\item The spin-independent scattering cross section is significant and a signal may appear with the next update from LUX and would be of the order of hundreds of events at Xenon-1T.
\item The spin-dependent scattering cross section puts the scenario within reach of IceCube after several years of operation. 
\end{itemize}
It is clear from this that should the GCE find its origin in the CMSSM the phenomenology would be extremely rich and striking. 

\begin{center}
\textbf{ACKNOWLEDGMENTS}
\end{center}

  We would like to thank Leszek Roszkowski and Enrico Sessolo for helpful comments and discussion.
  This work has been funded in part by the Welcome Programme
  of the Foundation for Polish Science.
  The use of the CIS computer cluster at the National Centre for Nuclear Research is gratefully acknowledged.

\bibliographystyle{utphysmcite}	

\bibliography{BF_13}

\end{document}

%% file: macros_BF_13.tex
\newcommand{\newc}{\newcommand*}

\long\def\begincomment#1\endcomment{%
        \begingroup\sf\baselineskip12pt#1\endgroup}

\newc{\etal}{\textrm{et al.}} 
\newc{\eg}{\textrm{e.g.}} 
\newc{\ie}{\textrm{i.e.}}
\newc{\etc}{\textrm{etc}}
\newc\vs{\textrm{vs.}}
\newc{\cl}{\rm {C.L.}}

\newc{\ev}{\ensuremath{\,\mathrm{eV}}}
\newc{\kev}{\ensuremath{\,\mathrm{keV}}}
\newc{\mev}{\ensuremath{\,\mathrm{MeV}}}
\newc{\gev}{\ensuremath{\,\mathrm{GeV}}}
\newc{\tev}{\ensuremath{\,\mathrm{TeV}}}
\newc{\MeV}{\mev} 
\newc{\TeV}{\tev}
\newc{\invpb}{\ensuremath{/\text{pb}}}
\newc{\invfb}{\ensuremath{\,\text{fb}^{-1}}}
\newc\nb{\ensuremath{\,\mathrm{nb}}} \newc\pb{\ensuremath{\,\mathrm{pb}}} \newc\fb{\ensuremath{\,\mathrm{fb}}}
\newc\pc{\ensuremath{\,\mathrm{pc}}}
\newc\kpc{\ensuremath{\,\mathrm{kpc}}}
\newc\mpc{\ensuremath{\,\mathrm{Mpc}}}
\newc\ps{\ensuremath{\,\mathrm{ps}}} 
\newc\cmeter{\ensuremath{\,\mathrm{cm}}} 
\newc\meter{\ensuremath{\,\mathrm{m}}} 
\newc\kmeter{\ensuremath{\,\mathrm{km}}}
\newc\second{\ensuremath{\,\mathrm{s}}}
\newc\msecond{\ensuremath{\,\mathrm{ms}}}
\newc\nsecond{\ensuremath{\,\mathrm{ns}}}
\newc\psecond{\ensuremath{\,\mathrm{ps}}}

\newc{\chisqmin}{\ensuremath{\chi^2_{\mathrm{min}}}}
\newc{\Delchisq}{\ensuremath{\Delta\chi^2}}
\newc{\chisq}{\ensuremath{\chi^2}}
\newc{\like}{\ensuremath{\mathcal{L}}}

\newc\lsim{\ensuremath{\mathrel{\rlap{\lower4pt\hbox{\hskip1pt$\sim$}}\raise1pt\hbox{$<$}}}}
\newc\gsim{\ensuremath{\mathrel{\rlap{\lower4pt\hbox{\hskip1pt$\sim$}}\raise1pt\hbox{$>$}}}}
\newc{\VEV}[1]{\ensuremath{\langle #1 \rangle}}
\newc{\dl}{\ensuremath{\stackrel{\leftarrow}{D}}}
\newc{\dr}{\ensuremath{\stackrel{\rightarrow}{D}}}

\newc{\bcenter}{\begin{center}}    \newc{\ecenter}{\end{center}}
\newc{\bfl}{\begin{flushleft}}    \newc{\efl}{\end{flushleft}}
\newc{\bfr}{\begin{flushright}}    \newc{\efr}{\end{flushright}}

\newc{\bi}{\begin{itemize}}
\newc{\ei}{\end{itemize}}
\newc{\bed}{\begin{description}}
\newc{\eed}{\end{description}}
\newc{\ben}{\begin{enumerate}}
\newc{\een}{\end{enumerate}}

\newc{\be}{\begin{equation}}
\newc{\ee}{\end{equation}}
\newc{\bea}{\begin{eqnarray}}
\newc{\eea}{\end{eqnarray}}
\newc{\ra}{\rightarrow}

\newc{\alphas}{\ensuremath{\alpha_s}}
\newc{\alphatwo}{\ensuremath{\alpha_2}}
\newc{\alphaone}{\ensuremath{\alpha_1}}
\newc{\alphai}[1]{\ensuremath{\alpha_{#1}}}
\newc{\alphaem}{\ensuremath{\alpha_{\mathrm{em}}}}
\newc{\alphaeff}{\ensuremath{\alpha_{\mathrm{eff}}}}
\newc{\sineff}{\ensuremath{\sin \theta_{\mathrm{eff}}}}
\newc{\sinsqeff}{\ensuremath{\sin^2 \theta_{\mathrm{eff}}}}
\newc{\dalphahad}{\ensuremath{\Delta \alpha_{\mathrm{had}}}}
\newc{\yt}{\ensuremath{h_t}} \newc{\yb}{\ensuremath{h_b}} \newc{\ytau}{\ensuremath{h_{\tau}}}
\newc\mz{\ensuremath{M_Z}} 
\newc\mw{\ensuremath{m_W}}
\newc\mZ{\mz}        \newc\mW{\mw}
\newc\mhsm{\ensuremath{ m_{H_{\mathrm{SM}}}}}
\newc{\mtop}{\ensuremath{ m_t}}               \newc{\mtpole}{\ensuremath{ M_t}}
\newc{\mbottom}{\ensuremath{ m_b}} 
\newc{\mtau}{\ensuremath{ m_{\tau}}}
\newc{\mt}{\mtpole}
\newc{\mb}{\mbottom} 
\newc{\rgg}{\ensuremath{R_{h}(\gamma\gamma)}}
\newc{\rzz}{\ensuremath{R_{h}(ZZ)}}
\newc{\rtwogg}{\ensuremath{R_{h_2}(\gamma\gamma)}}
\newc{\rtwozz}{\ensuremath{R_{h_2}(ZZ)}}
\newc{\ronegg}{\ensuremath{R_{h_1}(\gamma\gamma)}}
\newc{\ronezz}{\ensuremath{R_{h_1}(ZZ)}}
\newc{\rsiggg}{\ensuremath{R_{h_\textrm{sig}}(\gamma\gamma)}}
\newc{\rsigzz}{\ensuremath{R_{h_\textrm{sig}}(ZZ)}}
\newc{\llbar}{\ensuremath{\ell\bar{\ell}}}
\newc{\tauptaum}{\ensuremath{ \tau^+\tau^-}}
\newc{\qqbar}{\ensuremath{ q\bar{q}}} \newc{\ppbar}{\ensuremath{ p\bar{p}}}
\newc{\bbbar}{\ensuremath{ b\bar{b}}} \newc{\ttbar}{\ensuremath{ t\bar{t}}}
\newc{\ffbar}{\ensuremath{ f\bar{f}}} \newc{\tautaubar}{\ensuremath{ \tau\bar{\tau}}}
\newc{\mchi}{\ensuremath{m_{\chi}}}
\newc{\squark}{\ensuremath{\tilde{q}}}
\newc{\slepton}{\ensuremath{\tilde{l}}}
\newc{\gluino}{\ensuremath{\tilde{g}}} 
\newc{\mgluino}{\ensuremath{{m_{\gluino}}}}
\newc{\tone}{\ensuremath{{\tilde{t}_1}}}

\newc{\sthw}{\ensuremath{ \sin\theta_W}}              \newc{\cthw}{\ensuremath{\cos\theta_W}}
\newc{\tanthw}{\ensuremath{ \tan\theta_W}}              \newc{\cotthw}{\ensuremath{\cot\theta_W}}
\newc{\ssqthw}{\ensuremath{\sin^2 \theta_W}}
\newc{\msbar}{\ensuremath{\overline{MS}}} \newc{\drbar}{\ensuremath{\overline{DR}}}
\newc{\mtmtsmmsbar}{\ensuremath{ m_t(m_t)^{\msbar}_{{\mathrm{SM}}}}}
\newc{\mtmtsmdrbar}{\ensuremath{ m_t(m_t)^{\drbar}_{{\mathrm{SM}}}}}
\newc{\mtmtmssmdrbar}{\ensuremath{ m_t(m_t)^{\drbar}_{{\mathrm{SUSY}}}}}
\newc{\mbmbmsbar}{\ensuremath{ m_b(m_b)^{\msbar} }}
\newc{\mbmbsmmsbar}{\ensuremath{ m_b(m_b)^{\msbar}_{{\mathrm{SM}}}}}
\newc{\mbmzsmmsbar}{\ensuremath{ m_b(\mz)^{\msbar}_{{\mathrm{SM}}}}}
\newc{\mbmzsmdrbar}{\ensuremath{ m_b(\mz)^{\drbar}_{{\mathrm{SM}}}}}
\newc{\mbmzmssmdrbar}{\ensuremath{ m_b(\mz)^{\drbar}_{{\mathrm{SUSY}}}}}
\newc{\mtaumzsmmsbar}{\ensuremath{ m_{\tau}(\mz)^{\msbar}_{{\mathrm{SM}}}}}
\newc{\mtaumzsmdrbar}{\ensuremath{ m_{\tau}(\mz)^{\drbar}_{{\mathrm{SM}}}}}
\newc{\mtaumzmssmdrbar}{\ensuremath{ m_{\tau}(\mz)^{\drbar}_{{\mathrm{SUSY}}}}}
\newc{\alphasmzms}{\ensuremath{\alpha_s(M_Z)^{\overline{MS}}}}
\newc{\alphaimzms}[1]{\ensuremath{\alpha_{#1}(M_Z)^{\overline{MS}}}}
\newc{\alphaemmz}{\ensuremath{\alpha_{\mathrm{em}}(M_Z)^{\overline{MS}}}}

\newc{\mzero}{\ensuremath{{m_0}}}
\newc{\mhalf}{\ensuremath{ m_{1/2}}}
\newc{\tanb}{\ensuremath{\tan\beta}}
\newc{\azero}{\ensuremath{ A_0}}
\newc{\bzero}{\ensuremath{ B_0}}
\newc{\signmu}{\ensuremath{\rm{sgn}\,\mu}}
\newc{\mueff}{\ensuremath{\mu_{\rm{eff}}}}
\newc{\lam}{\ensuremath{{\lambda}}}
\newc{\kap}{\ensuremath{{\kappa}}}
\newc{\alam}{\ensuremath{{A_{\lambda}}}}
\newc{\akap}{\ensuremath{{A_{\kappa}}}}
\newc{\hs}{\ensuremath{ H_s}}      
\newc{\mhs}{\ensuremath{ m_{H_s}}} 
\newc{\mgut}{\ensuremath{ M_{\rm GUT}}}
\newc{\mplanck}{\ensuremath{ M_{\rm P}}}      \newc{\mpl}{\ensuremath{ M_{\rm Pl}}}
\newc{\msusy}{\ensuremath{ M_{\rm SUSY}}}      \newc{\ms}{\ensuremath{ M_{\rm S}}}
 \newc{\mhl}{\ensuremath{m_\hl}} 
 \newc{\mhone}{\ensuremath{m_{h_1}}} 
 \newc{\mhtwo}{\ensuremath{m_{h_2}}} 
 \newc{\mglu}{\ensuremath{m_{\tilde g}}} 
 \newc{\mul}{\ensuremath{m_{\tilde{u}_L}}} 
 \newc{\mtone}{\ensuremath{m_{\tilde{t}_1}}} 
 \newc{\ma}{\ensuremath{m_A}} 
 \newc{\maone}{\ensuremath{m_{a_1}}} 
 \newc{\matwo}{\ensuremath{m_{a_2}}}
 \newc{\hone}{\ensuremath{h_1}}
 \newc{\htwo}{\ensuremath{h_2}}
 \newc{\aone}{\ensuremath{a_1}}
 \newc{\atwo}{\ensuremath{a_2}}
 \newc{\mhu}{\ensuremath{ m_{H_u}}}       
 \newc{\mhd}{\ensuremath{ m_{H_d}}}
 \newc{\mhusq}{\ensuremath{ m_{H_u}^2}}       
 \newc{\mhdsq}{\ensuremath{ m_{H_d}^2}}
 \newc{\mhuew}{\ensuremath{ m^{\ast}_{H_u}}}       
 \newc{\mhdew}{\ensuremath{ m^{\ast}_{H_d}}}
 \newc{\mhuewsq}{\ensuremath{ m^{\ast\, 2}_{H_u}}}       
 \newc{\mhdewsq}{\ensuremath{ m^{\ast\, 2}_{H_d}}}
 \newc{\hu}{\ensuremath{ H_u}}       
 \newc{\hd}{\ensuremath{ H_d}}
 \newc{\barmhu}{\ensuremath{ \bar{m}_{H_u}}}
 \newc{\barmhd}{\ensuremath{ \bar{m}_{H_d}}}

 \newc{\mqthree}{\ensuremath{m_{\widetilde{Q}_3}^2}}
 \newc{\muthree}{\ensuremath{m_{\tilde{u}_3}^2}}
 \newc{\mdthree}{\ensuremath{m_{\tilde{d}_3}^2}}
 \newc{\mlthree}{\ensuremath{m_{\widetilde{L}_3}^2}}
 \newc{\methree}{\ensuremath{m_{\tilde{e}_3}^2}}
 \newc{\mqtwo}{\ensuremath{m_{\widetilde{Q}_2}^2}}
 \newc{\mutwo}{\ensuremath{m_{\tilde{u}_2}^2}}
 \newc{\mdtwo}{\ensuremath{m_{\tilde{d}_2}^2}}
 \newc{\mltwo}{\ensuremath{m_{\widetilde{L}_2}^2}}
 \newc{\metwo}{\ensuremath{m_{\tilde{e}_2}^2}}
 \newc{\mqone}{\ensuremath{m_{\widetilde{Q}_1}^2}}
 \newc{\muone}{\ensuremath{m_{\tilde{u}_1}^2}}
 \newc{\mdone}{\ensuremath{m_{\tilde{d}_1}^2}}
 \newc{\mlone}{\ensuremath{m_{\widetilde{L}_1}^2}}
 \newc{\meone}{\ensuremath{m_{\tilde{e}_1}^2}}
 \newc{\msmul}{\ensuremath{m_{\tilde{\mu}_L}}}
 \newc{\msmur}{\ensuremath{m_{\tilde{\mu}_R}}}
 \newc{\msneumu}{\ensuremath{m_{\tilde{\nu}_{\mu}}}}
 \newc{\mone}{\ensuremath{M_1}}
 \newc{\monesq}{\ensuremath{M_1^2}}
 \newc{\mtwo}{\ensuremath{M_2}}
 \newc{\mtwosq}{\ensuremath{M_2^2}}
 \newc{\mthree}{\ensuremath{M_3}}
 \newc{\mthreesq}{\ensuremath{M_3^2}}
 \newc{\atau}{\ensuremath{{A_{\tau}}}}
 \newc{\at}{\ensuremath{{A_{t}}}}
 \newc{\ab}{\ensuremath{{A_{b}}}}
 \newc{\atausq}{\ensuremath{{A_{\tau}^2}}}
 \newc{\atsq}{\ensuremath{{A_{t}^2}}}
 \newc{\absq}{\ensuremath{{A_{b}^2}}}

 \newc{\dmzero}{\ensuremath{\Delta{_{m_0}}}}
 \newc{\dmhalf}{\ensuremath{\Delta{_{m_{1/2}}}}}
 \newc{\dmu}{\ensuremath{\Delta{_{\mu}}}}

 \newc{\pten}{\ensuremath{\psi_{10}}}
 \newc{\ffive}{\ensuremath{\phi_{5}}}
 \newc{\hfive}{\ensuremath{h_{5}}}
 \newc{\hbfive}{\ensuremath{h_{\bar{5}}}}
 \newc{\thet}{\ensuremath{\theta_{50}}}
 \newc{\thetb}{\ensuremath{\theta_{\,\overline{50}}}}
 \newc{\ptenhat}{\ensuremath{\hat{\psi}_{10}}}
 \newc{\ffivehat}{\ensuremath{\hat{\phi}_{5}}}
 \newc{\hfivehat}{\ensuremath{\hat{h}_{5}}}
 \newc{\hbfivehat}{\ensuremath{\hat{h}_{\bar{5}}}}
 \newc{\thethat}{\ensuremath{\hat{\theta}_{50}}}
 \newc{\thetbhat}{\ensuremath{\hat{\theta}_{\,\overline{50}}}}
 \newc{\si}{\ensuremath{\Sigma}}
 \newc{\mfive}{\ensuremath{m_5^2}}
 \newc{\mten}{\ensuremath{m_{10}^2}}
 \newc{\dfive}{\ensuremath{\Delta^2_5}}
 \newc{\dbfive}{\ensuremath{\Delta^2_{\bar{5}}}}
 \newc{\dfifty}{\ensuremath{\Delta^2_{50}}}
 \newc{\dfiftyb}{\ensuremath{\Delta^2_{\,\overline{50}}}}
 \newc{\msi}{\ensuremath{m_{\Sigma}^2}}
 \newc{\lamh}{\ensuremath{\lambda_{H}}}
 \newc{\lamhb}{\ensuremath{\lambda_{\bar{H}}}}
 \newc{\ah}{\ensuremath{A_{H}}}
 \newc{\ahb}{\ensuremath{A_{\bar{H}}}}
 \newc{\lams}{\ensuremath{\lambda_{S}}}
 \newc{\as}{\ensuremath{A_{S}}}
 \newc{\lamsig}{\ensuremath{\lambda_{\si}}}
 \newc{\asig}{\ensuremath{A_{\si}}}

 \newc{\msten}{\ensuremath{m_{16}^2}}
 \newc{\mhun}{\ensuremath{m_{126}^2}}
 \newc{\mhunb}{\ensuremath{m_{\bar{126}}^2}}
 \newc{\mthun}{\ensuremath{m_{210}^2}}
 \newc{\ahun}{\ensuremath{A_{\bar{126}}}}
 \newc{\yhun}{\ensuremath{Y_{\bar{126}}}}
 \newc{\aten}{\ensuremath{A_{10}}}
 \newc{\yten}{\ensuremath{Y_{10}}}
 \newc{\alone}{\ensuremath{A_{\lambda_1}}}
 \newc{\altwo}{\ensuremath{A_{\lambda_2}}}
 \newc{\althree}{\ensuremath{A_{\lambda_3}}}
 \newc{\althreeb}{\ensuremath{A_{\bar{\lambda_3}}}}
 \newc{\lone}{\ensuremath{\lambda_1}}
 \newc{\ltwo}{\ensuremath{\lambda_2}}
 \newc{\lthree}{\ensuremath{\lambda_3}}
 \newc{\lthreeb}{\ensuremath{\bar{\lambda_3}}}

\newc{\sigsip}{\ensuremath{\sigma^{\rm SI}_{p}}}	\newc{\sigsin}{\ensuremath{\sigma^{\rm SI}_{n}}}
\newc{\sigsdp}{\ensuremath{\sigma^{\rm SD}_{p}}}	\newc{\sigsdn}{\ensuremath{\sigma^{\rm SD}_{n}}}
\newc{\sigsi}{\ensuremath{\sigma^{\rm SI}}}	\newc{\sigsd}{\ensuremath{\sigma^{\rm SD}}}
\newc{\sigv}{\ensuremath{\sigma v}}
\newc{\abund}{\ensuremath{ \Omega h^2}}
\newc{\omegadm}{\ensuremath{ \Omega_{{\rm DM}}}}     \newc{\abunddm}{\ensuremath{ \Omega_{{\rm DM}} h^2}} 
\newc{\omegam}{\ensuremath{ \Omega_{{\rm m}}}}       \newc{\abundm}{\ensuremath{ \Omega_{{\rm m}} h^2}}
\newc{\omegab}{\ensuremath{ \Omega_{{\rm b}}}}	\newc{\abundb}{\ensuremath{ \Omega_{{\rm b}} h^2}}
\newc{\omegatot}{\ensuremath{ \Omega_{{\rm TOT}}}}
\newc{\omegacdm}{\ensuremath{ \Omega_{{\rm CDM}}}}   \newc{\abundcdm}{\ensuremath{ \Omega_{{\rm CDM}} h^2}}
\newc{\omegalambda}{\ensuremath{ \Omega_{\Lambda}}} \newc{\abundlambda}{\ensuremath{ \Omega_{\Lambda} h^2}}
\newc{\omegarad}{\ensuremath{ \Omega_{{\rm rad}}}}  \newc{\abundrad}{\ensuremath{ \Omega_{{\rm rad}} h^2}}
\newc{\rhocrit}{\ensuremath{ \rho_{\rm crit}}}
\newc{\rhochi}{\ensuremath{ \rho_{\chi}}}
\newc{\abunchi}{\ensuremath{\Omega_\chi h^2}}
\newc{\abundlsp}{\ensuremath{\Omega_{\rm LSP}h^2}}
\newcommand*{\abundchi}{\ensuremath{\Omega_\chi h^2}}

\newc{\amu}{\ensuremath{ a_{\mu}}}        \newc{\amususy}{\ensuremath{ a_{\mu}^{\mathrm{SUSY}}}}
\newc{\amuexpt}{\ensuremath{ a_{\mu}^{\mathrm{expt}}}}        \newc{\amusm}{\ensuremath{ a_{\mu}^{\mathrm{SM}}}}
\newc\deltaamu{\ensuremath{\Delta a_{\mu}}} \newc{\deltaamususy}{\ensuremath{\delta a_{\mu}^{\mathrm{SUSY}}}}
\newc\gmtwo{\ensuremath{ (g-2)_{\mu}}} 
\newc{\deltagmtwomususy}{\ensuremath{\delta\left(g-2\right)_{\mu}^{\mathrm{SUSY}}}}
\newc{\deltagmtwomu}{\ensuremath{\delta\left(g-2\right)_{\mu}}}
\newc\BR{\ensuremath{\rm BR}}
\newc\bsgamma{\ensuremath{ b\rightarrow s \gamma }}
\newc\bxsgamma{\ensuremath{\overline{B}\rightarrow X_{s}\gamma}}
\newc\brbsgamma{\ensuremath{\BR\left(\bsgamma\right)}}
\newc\brbxsgamma{\ensuremath{\BR\left(\bxsgamma\right)}}
\newc\bsmumu{\ensuremath{B_s\to\mu^+\mu^-}}
\newc\brbsmumu{\ensuremath{\BR\left(B_s\to\mu^+\mu^-\right)}}
\newc\bdmmumu{\ensuremath{\overline{B}_d\to\mu^+\mu^-}}
\newc\bbbarmix{\ensuremath{\overline{B}_s\mbox{-}B_s}}      
\newc\delmbs{\ensuremath{\Delta M_{B_s}}}
\newc{\butaunu}{\ensuremath{B_u \rightarrow \tau \nu}}
\newc{\brbutaunu}{\ensuremath{\BR\left(B_u \rightarrow \tau \nu\right)}}

\newcommand*{\reffig}[1]{Fig.~\ref{#1}}

     \newcommand*{\refsec}[1]{Sec.~\ref{#1}}

\newcommand*{\feynhiggs}{FeynHiggs}
\newcommand*{\higgsbounds}{H{\scriptsize IGGS}B{\scriptsize OUNDS}}
\newcommand*{\higgssignals}{H{\scriptsize IGGS}S{\scriptsize IGNALS}}

\let\oldcite\cite
\renewcommand*{\cite}{~\oldcite}

\newcommand*{\hl}{\ensuremath{h}}



%% file: BF13_arxiv_v1.bbl
\ifx\mcitethebibliography\mciteundefinedmacro
\PackageError{unsrtM.bst}{mciteplus.sty has not been loaded}
{This bibstyle requires the use of the mciteplus package.}\fi
\begin{mcitethebibliography}{10}

\bibitem{Goodenough:2009gk}
L.~Goodenough and D.~Hooper, ``{Possible Evidence For Dark Matter Annihilation
  In The Inner Milky Way From The Fermi Gamma Ray Space Telescope},''
\href{http://arxiv.org/abs/0910.2998}{{\ttfamily arXiv:0910.2998 [hep-ph]}}.
\mciteBstWouldAddEndPunctfalse
\mciteSetBstMidEndSepPunct{\mcitedefaultmidpunct}
{}{\mcitedefaultseppunct}\relax
\EndOfBibitem
\bibitem{Boyarsky:2010dr}
A.~Boyarsky, D.~Malyshev, and O.~Ruchayskiy, ``{A comment on the emission from
  the Galactic Center as seen by the Fermi telescope},''
  \href{http://dx.doi.org/10.1016/j.physletb.2011.10.014}{{\em Phys. Lett.}
  {\bfseries B705} (2011) 165--169},
\href{http://arxiv.org/abs/1012.5839}{{\ttfamily arXiv:1012.5839 [hep-ph]}}.
\mciteBstWouldAddEndPunctfalse
\mciteSetBstMidEndSepPunct{\mcitedefaultmidpunct}
{}{\mcitedefaultseppunct}\relax
\EndOfBibitem
\bibitem{Hooper:2010mq}
D.~Hooper and L.~Goodenough, ``{Dark Matter Annihilation in The Galactic Center
  As Seen by the Fermi Gamma Ray Space Telescope},''
  \href{http://dx.doi.org/10.1016/j.physletb.2011.02.029}{{\em Phys. Lett.}
  {\bfseries B697} (2011) 412--428},
\href{http://arxiv.org/abs/1010.2752}{{\ttfamily arXiv:1010.2752 [hep-ph]}}.
\mciteBstWouldAddEndPunctfalse
\mciteSetBstMidEndSepPunct{\mcitedefaultmidpunct}
{}{\mcitedefaultseppunct}\relax
\EndOfBibitem
\bibitem{Hooper:2011ti}
D.~Hooper and T.~Linden, ``{On The Origin Of The Gamma Rays From The Galactic
  Center},'' \href{http://dx.doi.org/10.1103/PhysRevD.84.123005}{{\em Phys.
  Rev.} {\bfseries D84} (2011) 123005},
\href{http://arxiv.org/abs/1110.0006}{{\ttfamily arXiv:1110.0006
  [astro-ph.HE]}}.
\mciteBstWouldAddEndPunctfalse
\mciteSetBstMidEndSepPunct{\mcitedefaultmidpunct}
{}{\mcitedefaultseppunct}\relax
\EndOfBibitem
\bibitem{Abazajian:2012pn}
K.~N. Abazajian and M.~Kaplinghat, ``{Detection of a Gamma-Ray Source in the
  Galactic Center Consistent with Extended Emission from Dark Matter
  Annihilation and Concentrated Astrophysical Emission},''
  \href{http://dx.doi.org/10.1103/PhysRevD.86.083511,
  10.1103/PhysRevD.87.129902}{{\em Phys. Rev.} {\bfseries D86} (2012) 083511},
  \href{http://arxiv.org/abs/1207.6047}{{\ttfamily arXiv:1207.6047
  [astro-ph.HE]}}.
[Erratum: Phys. Rev.D87,129902(2013)].
\mciteBstWouldAddEndPunctfalse
\mciteSetBstMidEndSepPunct{\mcitedefaultmidpunct}
{}{\mcitedefaultseppunct}\relax
\EndOfBibitem
\bibitem{Linden:2012iv}
T.~Linden, E.~Lovegrove, and S.~Profumo, ``{The Morphology of Hadronic Emission
  Models for the Gamma-Ray Source at the Galactic Center},''
  \href{http://dx.doi.org/10.1088/0004-637X/753/1/41}{{\em Astrophys. J.}
  {\bfseries 753} (2012) 41},
\href{http://arxiv.org/abs/1203.3539}{{\ttfamily arXiv:1203.3539
  [astro-ph.HE]}}.
\mciteBstWouldAddEndPunctfalse
\mciteSetBstMidEndSepPunct{\mcitedefaultmidpunct}
{}{\mcitedefaultseppunct}\relax
\EndOfBibitem
\bibitem{Gordon:2013vta}
C.~Gordon and O.~Macias, ``{Dark Matter and Pulsar Model Constraints from
  Galactic Center Fermi-LAT Gamma Ray Observations},''
  \href{http://dx.doi.org/10.1103/PhysRevD.88.083521,
  10.1103/PhysRevD.89.049901}{{\em Phys. Rev.} {\bfseries D88} no.~8, (2013)
  083521}, \href{http://arxiv.org/abs/1306.5725}{{\ttfamily arXiv:1306.5725
  [astro-ph.HE]}}.
[Erratum: Phys. Rev.D89,no.4,049901(2014)].
\mciteBstWouldAddEndPunctfalse
\mciteSetBstMidEndSepPunct{\mcitedefaultmidpunct}
{}{\mcitedefaultseppunct}\relax
\EndOfBibitem
\bibitem{Hooper:2013rwa}
D.~Hooper and T.~R. Slatyer, ``{Two Emission Mechanisms in the Fermi Bubbles: A
  Possible Signal of Annihilating Dark Matter},''
  \href{http://dx.doi.org/10.1016/j.dark.2013.06.003}{{\em Phys. Dark Univ.}
  {\bfseries 2} (2013) 118--138},
\href{http://arxiv.org/abs/1302.6589}{{\ttfamily arXiv:1302.6589
  [astro-ph.HE]}}.
\mciteBstWouldAddEndPunctfalse
\mciteSetBstMidEndSepPunct{\mcitedefaultmidpunct}
{}{\mcitedefaultseppunct}\relax
\EndOfBibitem
\bibitem{Abazajian:2014fta}
K.~N. Abazajian, N.~Canac, S.~Horiuchi, and M.~Kaplinghat, ``{Astrophysical and
  Dark Matter Interpretations of Extended Gamma-Ray Emission from the Galactic
  Center},'' \href{http://dx.doi.org/10.1103/PhysRevD.90.023526}{{\em Phys.
  Rev.} {\bfseries D90} no.~2, (2014) 023526},
\href{http://arxiv.org/abs/1402.4090}{{\ttfamily arXiv:1402.4090
  [astro-ph.HE]}}.
\mciteBstWouldAddEndPunctfalse
\mciteSetBstMidEndSepPunct{\mcitedefaultmidpunct}
{}{\mcitedefaultseppunct}\relax
\EndOfBibitem
\bibitem{Daylan:2014rsa}
T.~Daylan, D.~P. Finkbeiner, D.~Hooper, T.~Linden, S.~K.~N. Portillo, N.~L.
  Rodd, and T.~R. Slatyer, ``{The Characterization of the Gamma-Ray Signal from
  the Central Milky Way: A Compelling Case for Annihilating Dark Matter},''
\href{http://arxiv.org/abs/1402.6703}{{\ttfamily arXiv:1402.6703
  [astro-ph.HE]}}.
\mciteBstWouldAddEndPunctfalse
\mciteSetBstMidEndSepPunct{\mcitedefaultmidpunct}
{}{\mcitedefaultseppunct}\relax
\EndOfBibitem
\bibitem{Calore:2014xka}
F.~Calore, I.~Cholis, and C.~Weniger, ``{Background model systematics for the
  Fermi GeV excess},''
  \href{http://dx.doi.org/10.1088/1475-7516/2015/03/038}{{\em JCAP} {\bfseries
  1503} (2015) 038},
\href{http://arxiv.org/abs/1409.0042}{{\ttfamily arXiv:1409.0042
  [astro-ph.CO]}}.
\mciteBstWouldAddEndPunctfalse
\mciteSetBstMidEndSepPunct{\mcitedefaultmidpunct}
{}{\mcitedefaultseppunct}\relax
\EndOfBibitem
\bibitem{Zhou:2014lva}
B.~Zhou, Y.-F. Liang, X.~Huang, X.~Li, Y.-Z. Fan, L.~Feng, and J.~Chang, ``{GeV
  excess in the Milky Way: The role of diffuse galactic gamma-ray emission
  templates},'' \href{http://dx.doi.org/10.1103/PhysRevD.91.123010}{{\em Phys.
  Rev.} {\bfseries D91} no.~12, (2015) 123010},
\href{http://arxiv.org/abs/1406.6948}{{\ttfamily arXiv:1406.6948
  [astro-ph.HE]}}.
\mciteBstWouldAddEndPunctfalse
\mciteSetBstMidEndSepPunct{\mcitedefaultmidpunct}
{}{\mcitedefaultseppunct}\relax
\EndOfBibitem
\bibitem{Murgia2014web}
S.~M. Fermi-LAT~Collaboration, ``Observation of the high energy gamma-ray
  emission towards the galactic center..''
  \url{http://fermi.gsfc.nasa.gov/science/mtgs/symposia/2014/program/08_Murgia.pdf},
  October, 2014\relax
\mciteBstWouldAddEndPunctfalse
\mciteSetBstMidEndSepPunct{\mcitedefaultmidpunct}
{}{\mcitedefaultseppunct}\relax
\EndOfBibitem
\bibitem{Abazajian:2010zy}
K.~N. Abazajian, ``{The Consistency of Fermi-LAT Observations of the Galactic
  Center with a Millisecond Pulsar Population in the Central Stellar
  Cluster},'' \href{http://dx.doi.org/10.1088/1475-7516/2011/03/010}{{\em JCAP}
  {\bfseries 1103} (2011) 010},
\href{http://arxiv.org/abs/1011.4275}{{\ttfamily arXiv:1011.4275
  [astro-ph.HE]}}.
\mciteBstWouldAddEndPunctfalse
\mciteSetBstMidEndSepPunct{\mcitedefaultmidpunct}
{}{\mcitedefaultseppunct}\relax
\EndOfBibitem
\bibitem{Calore:2014oga}
F.~Calore, M.~Di~Mauro, F.~Donato, and F.~Donato, ``{Diffuse gamma-ray emission
  from galactic pulsars},''
  \href{http://dx.doi.org/10.1088/0004-637X/796/1/14}{{\em Astrophys. J.}
  {\bfseries 796} (2014) 1},
\href{http://arxiv.org/abs/1406.2706}{{\ttfamily arXiv:1406.2706
  [astro-ph.HE]}}.
\mciteBstWouldAddEndPunctfalse
\mciteSetBstMidEndSepPunct{\mcitedefaultmidpunct}
{}{\mcitedefaultseppunct}\relax
\EndOfBibitem
\bibitem{YusefZadeh:2012nh}
F.~Yusef-Zadeh {\em et~al.}, ``{Interacting Cosmic Rays with Molecular Clouds:
  A Bremsstrahlung Origin of Diffuse High Energy Emission from the Inner 2deg
  by 1deg of the Galactic Center},''
  \href{http://dx.doi.org/10.1088/0004-637X/762/1/33}{{\em Astrophys. J.}
  {\bfseries 762} (2013) 33},
\href{http://arxiv.org/abs/1206.6882}{{\ttfamily arXiv:1206.6882
  [astro-ph.HE]}}.
\mciteBstWouldAddEndPunctfalse
\mciteSetBstMidEndSepPunct{\mcitedefaultmidpunct}
{}{\mcitedefaultseppunct}\relax
\EndOfBibitem
\bibitem{Bartels:2015aea}
R.~Bartels, S.~Krishnamurthy, and C.~Weniger, ``{Strong support for the
  millisecond pulsar origin of the Galactic center GeV excess},''
\href{http://arxiv.org/abs/1506.05104}{{\ttfamily arXiv:1506.05104
  [astro-ph.HE]}}.
\mciteBstWouldAddEndPunctfalse
\mciteSetBstMidEndSepPunct{\mcitedefaultmidpunct}
{}{\mcitedefaultseppunct}\relax
\EndOfBibitem
\bibitem{Lee:2015fea}
S.~K. Lee, M.~Lisanti, B.~R. Safdi, T.~R. Slatyer, and W.~Xue, ``{Evidence for
  Unresolved Gamma-Ray Point Sources in the Inner Galaxy},''
\href{http://arxiv.org/abs/1506.05124}{{\ttfamily arXiv:1506.05124
  [astro-ph.HE]}}.
\mciteBstWouldAddEndPunctfalse
\mciteSetBstMidEndSepPunct{\mcitedefaultmidpunct}
{}{\mcitedefaultseppunct}\relax
\EndOfBibitem
\bibitem{Geringer-Sameth:2015lua}
A.~Geringer-Sameth, M.~G. Walker, S.~M. Koushiappas, S.~E. Koposov,
  V.~Belokurov, G.~Torrealba, and N.~W. Evans, ``{Indication of Gamma-ray
  Emission from the Newly Discovered Dwarf Galaxy Reticulum II},''
  \href{http://dx.doi.org/10.1103/PhysRevLett.115.081101}{{\em Phys. Rev.
  Lett.} {\bfseries 115} no.~8, (2015) 081101},
\href{http://arxiv.org/abs/1503.02320}{{\ttfamily arXiv:1503.02320
  [astro-ph.HE]}}.
\mciteBstWouldAddEndPunctfalse
\mciteSetBstMidEndSepPunct{\mcitedefaultmidpunct}
{}{\mcitedefaultseppunct}\relax
\EndOfBibitem
\bibitem{Caron:2015wda}
A.~Achterberg, S.~Amoroso, S.~Caron, L.~Hendriks, R.~Ruiz~de Austri, and
  C.~Weniger, ``{A description of the Galactic Center excess in the Minimal
  Supersymmetric Standard Model},''
  \href{http://dx.doi.org/10.1088/1475-7516/2015/08/006}{{\em JCAP} {\bfseries
  1508} no.~08, (2015) 006},
\href{http://arxiv.org/abs/1502.05703}{{\ttfamily arXiv:1502.05703 [hep-ph]}}.
\mciteBstWouldAddEndPunctfalse
\mciteSetBstMidEndSepPunct{\mcitedefaultmidpunct}
{}{\mcitedefaultseppunct}\relax
\EndOfBibitem
\bibitem{Bertone:2015tza}
G.~Bertone, F.~Calore, S.~Caron, R.~R. de~Austri, J.~S. Kim, R.~Trotta, and
  C.~Weniger, ``{Global analysis of the pMSSM in light of the Fermi GeV excess:
  prospects for the LHC Run-II and astroparticle experiments},''
\href{http://arxiv.org/abs/1507.07008}{{\ttfamily arXiv:1507.07008 [hep-ph]}}.
\mciteBstWouldAddEndPunctfalse
\mciteSetBstMidEndSepPunct{\mcitedefaultmidpunct}
{}{\mcitedefaultseppunct}\relax
\EndOfBibitem
\bibitem{Freese:2015ysa}
K.~Freese, A.~Lopez, N.~R. Shah, and B.~Shakya, ``{MSSM A-funnel and the
  Galactic Center Excess: Prospects for the LHC and Direct Detection
  Experiments},''
\href{http://arxiv.org/abs/1509.05076}{{\ttfamily arXiv:1509.05076 [hep-ph]}}.
\mciteBstWouldAddEndPunctfalse
\mciteSetBstMidEndSepPunct{\mcitedefaultmidpunct}
{}{\mcitedefaultseppunct}\relax
\EndOfBibitem
\bibitem{Bi:2015qva}
X.-J. Bi, L.~Bian, W.~Huang, J.~Shu, and P.-F. Yin, ``{Interpretation of the
  Galactic Center excess and electroweak phase transition in the NMSSM},''
  \href{http://dx.doi.org/10.1103/PhysRevD.92.023507}{{\em Phys. Rev.}
  {\bfseries D92} no.~2, (2015) 023507},
\href{http://arxiv.org/abs/1503.03749}{{\ttfamily arXiv:1503.03749 [hep-ph]}}.
\mciteBstWouldAddEndPunctfalse
\mciteSetBstMidEndSepPunct{\mcitedefaultmidpunct}
{}{\mcitedefaultseppunct}\relax
\EndOfBibitem
\bibitem{Cao:2015loa}
J.~Cao, L.~Shang, P.~Wu, J.~M. Yang, and Y.~Zhang, ``{Interpreting the galactic
  center gamma-ray excess in the NMSSM},''
\href{http://arxiv.org/abs/1506.06471}{{\ttfamily arXiv:1506.06471 [hep-ph]}}.
\mciteBstWouldAddEndPunctfalse
\mciteSetBstMidEndSepPunct{\mcitedefaultmidpunct}
{}{\mcitedefaultseppunct}\relax
\EndOfBibitem
\bibitem{Cheung:2014lqa}
C.~Cheung, M.~Papucci, D.~Sanford, N.~R. Shah, and K.~M. Zurek, ``{NMSSM
  Interpretation of the Galactic Center Excess},''
  \href{http://dx.doi.org/10.1103/PhysRevD.90.075011}{{\em Phys. Rev.}
  {\bfseries D90} no.~7, (2014) 075011},
\href{http://arxiv.org/abs/1406.6372}{{\ttfamily arXiv:1406.6372 [hep-ph]}}.
\mciteBstWouldAddEndPunctfalse
\mciteSetBstMidEndSepPunct{\mcitedefaultmidpunct}
{}{\mcitedefaultseppunct}\relax
\EndOfBibitem
\bibitem{Guo:2014gra}
J.~Guo, J.~Li, T.~Li, and A.~G. Williams, ``{NMSSM explanations of the Galactic
  center gamma ray excess and promising LHC searches},''
  \href{http://dx.doi.org/10.1103/PhysRevD.91.095003}{{\em Phys. Rev.}
  {\bfseries D91} no.~9, (2015) 095003},
\href{http://arxiv.org/abs/1409.7864}{{\ttfamily arXiv:1409.7864 [hep-ph]}}.
\mciteBstWouldAddEndPunctfalse
\mciteSetBstMidEndSepPunct{\mcitedefaultmidpunct}
{}{\mcitedefaultseppunct}\relax
\EndOfBibitem
\bibitem{Cahill-Rowley:2014ora}
M.~Cahill-Rowley, J.~Gainer, J.~Hewett, and T.~Rizzo, ``{Towards a
  Supersymmetric Description of the Fermi Galactic Center Excess},''
  \href{http://dx.doi.org/10.1007/JHEP02(2015)057}{{\em JHEP} {\bfseries 02}
  (2015) 057},
\href{http://arxiv.org/abs/1409.1573}{{\ttfamily arXiv:1409.1573 [hep-ph]}}.
\mciteBstWouldAddEndPunctfalse
\mciteSetBstMidEndSepPunct{\mcitedefaultmidpunct}
{}{\mcitedefaultseppunct}\relax
\EndOfBibitem
\bibitem{Agrawal:2014oha}
P.~Agrawal, B.~Batell, P.~J. Fox, and R.~Harnik, ``{WIMPs at the Galactic
  Center},'' \href{http://dx.doi.org/10.1088/1475-7516/2015/05/011}{{\em JCAP}
  {\bfseries 1505} (2015) 011},
\href{http://arxiv.org/abs/1411.2592}{{\ttfamily arXiv:1411.2592 [hep-ph]}}.
\mciteBstWouldAddEndPunctfalse
\mciteSetBstMidEndSepPunct{\mcitedefaultmidpunct}
{}{\mcitedefaultseppunct}\relax
\EndOfBibitem
\bibitem{Kane:1993td}
G.~L. Kane, C.~F. Kolda, L.~Roszkowski, and J.~D. Wells, ``{Study of
  constrained minimal supersymmetry},''
  \href{http://dx.doi.org/10.1103/PhysRevD.49.6173}{{\em Phys.Rev.} {\bfseries
  D49} (1994) 6173--6210},
\href{http://arxiv.org/abs/hep-ph/9312272}{{\ttfamily arXiv:hep-ph/9312272
  [hep-ph]}}.
\mciteBstWouldAddEndPunctfalse
\mciteSetBstMidEndSepPunct{\mcitedefaultmidpunct}
{}{\mcitedefaultseppunct}\relax
\EndOfBibitem
\bibitem{Baer:2011ab}
H.~Baer, V.~Barger, and A.~Mustafayev, ``{Implications of a 125 GeV Higgs
  scalar for LHC SUSY and neutralino dark matter searches},''
  \href{http://dx.doi.org/10.1103/PhysRevD.85.075010}{{\em Phys.Rev.}
  {\bfseries D85} (2012) 075010},
\href{http://arxiv.org/abs/1112.3017}{{\ttfamily arXiv:1112.3017 [hep-ph]}}.
\mciteBstWouldAddEndPunctfalse
\mciteSetBstMidEndSepPunct{\mcitedefaultmidpunct}
{}{\mcitedefaultseppunct}\relax
\EndOfBibitem
\bibitem{Baer:2012uya}
H.~Baer, V.~Barger, and A.~Mustafayev, ``{Neutralino dark matter in
  mSUGRA/CMSSM with a 125 GeV light Higgs scalar},''
  \href{http://dx.doi.org/10.1007/JHEP05(2012)091}{{\em JHEP} {\bfseries 1205}
  (2012) 091},
\href{http://arxiv.org/abs/1202.4038}{{\ttfamily arXiv:1202.4038 [hep-ph]}}.
\mciteBstWouldAddEndPunctfalse
\mciteSetBstMidEndSepPunct{\mcitedefaultmidpunct}
{}{\mcitedefaultseppunct}\relax
\EndOfBibitem
\bibitem{Kadastik:2011aa}
M.~Kadastik, K.~Kannike, A.~Racioppi, and M.~Raidal, ``{Implications of the 125
  GeV Higgs boson for scalar dark matter and for the CMSSM phenomenology},''
  \href{http://dx.doi.org/10.1007/JHEP05(2012)061}{{\em JHEP} {\bfseries 1205}
  (2012) 061},
\href{http://arxiv.org/abs/1112.3647}{{\ttfamily arXiv:1112.3647 [hep-ph]}}.
\mciteBstWouldAddEndPunctfalse
\mciteSetBstMidEndSepPunct{\mcitedefaultmidpunct}
{}{\mcitedefaultseppunct}\relax
\EndOfBibitem
\bibitem{Cao:2011sn}
J.~Cao, Z.~Heng, D.~Li, and J.~M. Yang, ``{Current experimental constraints on
  the lightest Higgs boson mass in the constrained MSSM},''
  \href{http://dx.doi.org/10.1016/j.physletb.2012.03.052}{{\em Phys.Lett.}
  {\bfseries B710} (2012) 665--670},
\href{http://arxiv.org/abs/1112.4391}{{\ttfamily arXiv:1112.4391 [hep-ph]}}.
\mciteBstWouldAddEndPunctfalse
\mciteSetBstMidEndSepPunct{\mcitedefaultmidpunct}
{}{\mcitedefaultseppunct}\relax
\EndOfBibitem
\bibitem{Ellis:2012aa}
J.~Ellis and K.~A. Olive, ``{Revisiting the Higgs Mass and Dark Matter in the
  CMSSM},'' \href{http://dx.doi.org/10.1140/epjc/s10052-012-2005-2}{{\em
  Eur.Phys.J.} {\bfseries C72} (2012) 2005},
\href{http://arxiv.org/abs/1202.3262}{{\ttfamily arXiv:1202.3262 [hep-ph]}}.
\mciteBstWouldAddEndPunctfalse
\mciteSetBstMidEndSepPunct{\mcitedefaultmidpunct}
{}{\mcitedefaultseppunct}\relax
\EndOfBibitem
\bibitem{Bechtle:2012zk}
P.~Bechtle, T.~Bringmann, K.~Desch, H.~Dreiner, M.~Hamer, {\em et~al.},
  ``{Constrained Supersymmetry after two years of LHC data: a global view with
  Fittino},'' \href{http://dx.doi.org/10.1007/JHEP06(2012)098}{{\em JHEP}
  {\bfseries 1206} (2012) 098},
\href{http://arxiv.org/abs/1204.4199}{{\ttfamily arXiv:1204.4199 [hep-ph]}}.
\mciteBstWouldAddEndPunctfalse
\mciteSetBstMidEndSepPunct{\mcitedefaultmidpunct}
{}{\mcitedefaultseppunct}\relax
\EndOfBibitem
\bibitem{Balazs:2013qva}
C.~Balazs, A.~Buckley, D.~Carter, B.~Farmer, and M.~White, ``{Should we still
  believe in constrained supersymmetry?},''
  \href{http://dx.doi.org/10.1140/epjc/s10052-013-2563-y}{{\em Eur.Phys.J.}
  {\bfseries C73} (2013) 2563},
\href{http://arxiv.org/abs/1205.1568}{{\ttfamily arXiv:1205.1568 [hep-ph]}}.
\mciteBstWouldAddEndPunctfalse
\mciteSetBstMidEndSepPunct{\mcitedefaultmidpunct}
{}{\mcitedefaultseppunct}\relax
\EndOfBibitem
\bibitem{Fowlie:2012im}
A.~Fowlie, M.~Kazana, K.~Kowalska, S.~Munir, L.~Roszkowski, {\em et~al.},
  ``{The CMSSM Favoring New Territories: The Impact of New LHC Limits and a 125
  GeV Higgs},'' \href{http://dx.doi.org/10.1103/PhysRevD.86.075010}{{\em
  Phys.Rev.} {\bfseries D86} (2012) 075010},
\href{http://arxiv.org/abs/1206.0264}{{\ttfamily arXiv:1206.0264 [hep-ph]}}.
\mciteBstWouldAddEndPunctfalse
\mciteSetBstMidEndSepPunct{\mcitedefaultmidpunct}
{}{\mcitedefaultseppunct}\relax
\EndOfBibitem
\bibitem{Akula:2012kk}
S.~Akula, P.~Nath, and G.~Peim, ``{Implications of the Higgs Boson Discovery
  for mSUGRA},'' \href{http://dx.doi.org/10.1016/j.physletb.2012.09.007}{{\em
  Phys.Lett.} {\bfseries B717} (2012) 188--192},
\href{http://arxiv.org/abs/1207.1839}{{\ttfamily arXiv:1207.1839 [hep-ph]}}.
\mciteBstWouldAddEndPunctfalse
\mciteSetBstMidEndSepPunct{\mcitedefaultmidpunct}
{}{\mcitedefaultseppunct}\relax
\EndOfBibitem
\bibitem{Buchmueller:2012hv}
O.~Buchmueller, R.~Cavanaugh, M.~Citron, A.~De~Roeck, M.~Dolan, {\em et~al.},
  ``{The CMSSM and NUHM1 in Light of 7 TeV LHC, $B_s$ to mu+mu- and XENON100
  Data},'' \href{http://dx.doi.org/10.1140/epjc/s10052-012-2243-3}{{\em
  Eur.Phys.J.} {\bfseries C72} (2012) 2243},
\href{http://arxiv.org/abs/1207.7315}{{\ttfamily arXiv:1207.7315}}.
\mciteBstWouldAddEndPunctfalse
\mciteSetBstMidEndSepPunct{\mcitedefaultmidpunct}
{}{\mcitedefaultseppunct}\relax
\EndOfBibitem
\bibitem{Strege:2012bt}
C.~Strege, G.~Bertone, F.~Feroz, M.~Fornasa, R.~Ruiz~de Austri, {\em et~al.},
  ``{Global Fits of the cMSSM and NUHM including the LHC Higgs discovery and
  new XENON100 constraints},''
  \href{http://dx.doi.org/10.1088/1475-7516/2013/04/013}{{\em JCAP} {\bfseries
  1304} (2013) 013},
\href{http://arxiv.org/abs/1212.2636}{{\ttfamily arXiv:1212.2636 [hep-ph]}}.
\mciteBstWouldAddEndPunctfalse
\mciteSetBstMidEndSepPunct{\mcitedefaultmidpunct}
{}{\mcitedefaultseppunct}\relax
\EndOfBibitem
\bibitem{Cabrera:2012vu}
M.~E. Cabrera, J.~A. Casas, and R.~R. de~Austri, ``{The health of SUSY after
  the Higgs discovery and the XENON100 data},''
  \href{http://dx.doi.org/10.1007/JHEP07(2013)182}{{\em JHEP} {\bfseries 1307}
  (2013) 182},
\href{http://arxiv.org/abs/1212.4821}{{\ttfamily arXiv:1212.4821 [hep-ph]}}.
\mciteBstWouldAddEndPunctfalse
\mciteSetBstMidEndSepPunct{\mcitedefaultmidpunct}
{}{\mcitedefaultseppunct}\relax
\EndOfBibitem
\bibitem{Kowalska:2013hha}
K.~Kowalska, L.~Roszkowski, and E.~M. Sessolo, ``{Two ultimate tests of
  constrained supersymmetry},''
  \href{http://dx.doi.org/10.1007/JHEP06(2013)078}{{\em JHEP} {\bfseries 1306}
  (2013) 078},
\href{http://arxiv.org/abs/1302.5956}{{\ttfamily arXiv:1302.5956 [hep-ph]}}.
\mciteBstWouldAddEndPunctfalse
\mciteSetBstMidEndSepPunct{\mcitedefaultmidpunct}
{}{\mcitedefaultseppunct}\relax
\EndOfBibitem
\bibitem{Dighe:2013wfa}
A.~Dighe, D.~Ghosh, K.~M. Patel, and S.~Raychaudhuri, ``{Testing Times for
  Supersymmetry: Looking Under the Lamp Post},''
  \href{http://dx.doi.org/10.1142/S0217751X13501340}{{\em Int.J.Mod.Phys.}
  {\bfseries A28} (2013) 1350134},
\href{http://arxiv.org/abs/1303.0721}{{\ttfamily arXiv:1303.0721 [hep-ph]}}.
\mciteBstWouldAddEndPunctfalse
\mciteSetBstMidEndSepPunct{\mcitedefaultmidpunct}
{}{\mcitedefaultseppunct}\relax
\EndOfBibitem
\bibitem{Buchmueller:2013rsa}
O.~Buchmueller {\em et~al.}, ``{The CMSSM and NUHM1 after LHC Run 1},''
  \href{http://dx.doi.org/10.1140/epjc/s10052-014-2922-3}{{\em Eur. Phys. J.}
  {\bfseries C74} no.~6, (2014) 2922},
\href{http://arxiv.org/abs/1312.5250}{{\ttfamily arXiv:1312.5250 [hep-ph]}}.
\mciteBstWouldAddEndPunctfalse
\mciteSetBstMidEndSepPunct{\mcitedefaultmidpunct}
{}{\mcitedefaultseppunct}\relax
\EndOfBibitem
\bibitem{Roszkowski:2014wqa}
L.~Roszkowski, E.~M. Sessolo, and A.~J. Williams, ``{What next for the CMSSM
  and the NUHM: Improved prospects for superpartner and dark matter
  detection},'' \href{http://dx.doi.org/10.1007/JHEP08(2014)067}{{\em JHEP}
  {\bfseries 08} (2014) 067},
\href{http://arxiv.org/abs/1405.4289}{{\ttfamily arXiv:1405.4289 [hep-ph]}}.
\mciteBstWouldAddEndPunctfalse
\mciteSetBstMidEndSepPunct{\mcitedefaultmidpunct}
{}{\mcitedefaultseppunct}\relax
\EndOfBibitem
\bibitem{Bechtle:2015nua}
P.~Bechtle {\em et~al.}, ``{Killing the cMSSM softly},''
\href{http://arxiv.org/abs/1508.05951}{{\ttfamily arXiv:1508.05951 [hep-ph]}}.
\mciteBstWouldAddEndPunctfalse
\mciteSetBstMidEndSepPunct{\mcitedefaultmidpunct}
{}{\mcitedefaultseppunct}\relax
\EndOfBibitem
\bibitem{Fowlie:2011mb}
A.~Fowlie, A.~Kalinowski, M.~Kazana, L.~Roszkowski, and Y.~S. Tsai, ``{Bayesian
  Implications of Current LHC and XENON100 Search Limits for the Constrained
  MSSM},'' \href{http://dx.doi.org/10.1103/PhysRevD.85.075012}{{\em Phys.Rev.}
  {\bfseries D85} (2012) 075012},
\href{http://arxiv.org/abs/1111.6098}{{\ttfamily arXiv:1111.6098 [hep-ph]}}.
\mciteBstWouldAddEndPunctfalse
\mciteSetBstMidEndSepPunct{\mcitedefaultmidpunct}
{}{\mcitedefaultseppunct}\relax
\EndOfBibitem
\bibitem{Roszkowski:2012uf}
L.~Roszkowski, E.~M. Sessolo, and Y.-L.~S. Tsai, ``{Bayesian Implications of
  Current LHC Supersymmetry and Dark Matter Detection Searches for the
  Constrained MSSM},'' \href{http://dx.doi.org/10.1103/PhysRevD.86.095005}{{\em
  Phys.Rev.} {\bfseries D86} (2012) 095005},
\href{http://arxiv.org/abs/1202.1503}{{\ttfamily arXiv:1202.1503 [hep-ph]}}.
\mciteBstWouldAddEndPunctfalse
\mciteSetBstMidEndSepPunct{\mcitedefaultmidpunct}
{}{\mcitedefaultseppunct}\relax
\EndOfBibitem
\bibitem{deAustri:2006pe}
R.~R. de~Austri, R.~Trotta, and L.~Roszkowski, ``{A Markov chain Monte Carlo
  analysis of the CMSSM},''
  \href{http://dx.doi.org/10.1088/1126-6708/2006/05/002}{{\em JHEP} {\bfseries
  05} (2006) 002},
\href{http://arxiv.org/abs/hep-ph/0602028}{{\ttfamily arXiv:hep-ph/0602028
  [hep-ph]}}.
\mciteBstWouldAddEndPunctfalse
\mciteSetBstMidEndSepPunct{\mcitedefaultmidpunct}
{}{\mcitedefaultseppunct}\relax
\EndOfBibitem
\bibitem{Bechtle:2013xfa}
P.~Bechtle, S.~Heinemeyer, O.~Stål, T.~Stefaniak, and G.~Weiglein,
  ``{$HiggsSignals$: Confronting arbitrary Higgs sectors with measurements at
  the Tevatron and the LHC},''
  \href{http://dx.doi.org/10.1140/epjc/s10052-013-2711-4}{{\em Eur. Phys. J.}
  {\bfseries C74} no.~2, (2014) 2711},
\href{http://arxiv.org/abs/1305.1933}{{\ttfamily arXiv:1305.1933 [hep-ph]}}.
\mciteBstWouldAddEndPunctfalse
\mciteSetBstMidEndSepPunct{\mcitedefaultmidpunct}
{}{\mcitedefaultseppunct}\relax
\EndOfBibitem
\bibitem{Bechtle:2008jh}
P.~Bechtle, O.~Brein, S.~Heinemeyer, G.~Weiglein, and K.~E. Williams,
  ``{HiggsBounds: Confronting Arbitrary Higgs Sectors with Exclusion Bounds
  from LEP and the Tevatron},''
  \href{http://dx.doi.org/10.1016/j.cpc.2009.09.003}{{\em Comput.Phys.Commun.}
  {\bfseries 181} (2010) 138--167},
\href{http://arxiv.org/abs/0811.4169}{{\ttfamily arXiv:0811.4169 [hep-ph]}}.
\mciteBstWouldAddEndPunctfalse
\mciteSetBstMidEndSepPunct{\mcitedefaultmidpunct}
{}{\mcitedefaultseppunct}\relax
\EndOfBibitem
\bibitem{Bechtle:2011sb}
P.~Bechtle, O.~Brein, S.~Heinemeyer, G.~Weiglein, and K.~E. Williams,
  ``{HiggsBounds 2.0.0: Confronting Neutral and Charged Higgs Sector
  Predictions with Exclusion Bounds from LEP and the Tevatron},''
  \href{http://dx.doi.org/10.1016/j.cpc.2011.07.015}{{\em Comput.Phys.Commun.}
  {\bfseries 182} (2011) 2605--2631},
\href{http://arxiv.org/abs/1102.1898}{{\ttfamily arXiv:1102.1898 [hep-ph]}}.
\mciteBstWouldAddEndPunctfalse
\mciteSetBstMidEndSepPunct{\mcitedefaultmidpunct}
{}{\mcitedefaultseppunct}\relax
\EndOfBibitem
\bibitem{Bechtle:2013wla}
P.~Bechtle, O.~Brein, S.~Heinemeyer, O.~St{\aa}l, T.~Stefaniak, {\em et~al.},
  ``{HiggsBounds-4: Improved Tests of Extended Higgs Sectors against Exclusion
  Bounds from LEP, the Tevatron and the LHC},''
  \href{http://dx.doi.org/10.1140/epjc/s10052-013-2693-2}{{\em Eur.Phys.J.}
  {\bfseries C74} (2014) 2693},
\href{http://arxiv.org/abs/1311.0055}{{\ttfamily arXiv:1311.0055 [hep-ph]}}.
\mciteBstWouldAddEndPunctfalse
\mciteSetBstMidEndSepPunct{\mcitedefaultmidpunct}
{}{\mcitedefaultseppunct}\relax
\EndOfBibitem
\bibitem{Heinemeyer:1998np}
S.~Heinemeyer, W.~Hollik, and G.~Weiglein, ``{The Masses of the neutral CP -
  even Higgs bosons in the MSSM: Accurate analysis at the two loop level},''
  \href{http://dx.doi.org/10.1007/s100529900006}{{\em Eur.Phys.J.} {\bfseries
  C9} (1999) 343--366},
\href{http://arxiv.org/abs/hep-ph/9812472}{{\ttfamily arXiv:hep-ph/9812472
  [hep-ph]}}.
\mciteBstWouldAddEndPunctfalse
\mciteSetBstMidEndSepPunct{\mcitedefaultmidpunct}
{}{\mcitedefaultseppunct}\relax
\EndOfBibitem
\bibitem{Akerib:2013tjd}
{\bfseries LUX} Collaboration, D.~S. Akerib {\em et~al.}, ``{First results from
  the LUX dark matter experiment at the Sanford Underground Research
  Facility},'' \href{http://dx.doi.org/10.1103/PhysRevLett.112.091303}{{\em
  Phys. Rev. Lett.} {\bfseries 112} (2014) 091303},
\href{http://arxiv.org/abs/1310.8214}{{\ttfamily arXiv:1310.8214
  [astro-ph.CO]}}.
\mciteBstWouldAddEndPunctfalse
\mciteSetBstMidEndSepPunct{\mcitedefaultmidpunct}
{}{\mcitedefaultseppunct}\relax
\EndOfBibitem
\bibitem{Kowalska:2014hza}
K.~Kowalska, L.~Roszkowski, E.~M. Sessolo, and S.~Trojanowski, ``{Low fine
  tuning in the MSSM with higgsino dark matter and unification constraints},''
  \href{http://dx.doi.org/10.1007/JHEP04(2014)166}{{\em JHEP} {\bfseries 04}
  (2014) 166},
\href{http://arxiv.org/abs/1402.1328}{{\ttfamily arXiv:1402.1328 [hep-ph]}}.
\mciteBstWouldAddEndPunctfalse
\mciteSetBstMidEndSepPunct{\mcitedefaultmidpunct}
{}{\mcitedefaultseppunct}\relax
\EndOfBibitem
\bibitem{Aartsen:2012kia}
{\bfseries IceCube} Collaboration, M.~G. Aartsen {\em et~al.}, ``{Search for
  dark matter annihilations in the Sun with the 79-string IceCube detector},''
  \href{http://dx.doi.org/10.1103/PhysRevLett.110.131302}{{\em Phys. Rev.
  Lett.} {\bfseries 110} no.~13, (2013) 131302},
\href{http://arxiv.org/abs/1212.4097}{{\ttfamily arXiv:1212.4097
  [astro-ph.HE]}}.
\mciteBstWouldAddEndPunctfalse
\mciteSetBstMidEndSepPunct{\mcitedefaultmidpunct}
{}{\mcitedefaultseppunct}\relax
\EndOfBibitem
\bibitem{Ackermann:2015zua}
{\bfseries Fermi-LAT} Collaboration, M.~Ackermann {\em et~al.}, ``{Searching
  for Dark Matter Annihilation from Milky Way Dwarf Spheroidal Galaxies with
  Six Years of Fermi-LAT Data},''
\href{http://arxiv.org/abs/1503.02641}{{\ttfamily arXiv:1503.02641
  [astro-ph.HE]}}.
\mciteBstWouldAddEndPunctfalse
\mciteSetBstMidEndSepPunct{\mcitedefaultmidpunct}
{}{\mcitedefaultseppunct}\relax
\EndOfBibitem
\bibitem{Ade:2013zuv}
{\bfseries Planck} Collaboration, P.~A.~R. Ade {\em et~al.}, ``{Planck 2013
  results. XVI. Cosmological parameters},''
  \href{http://dx.doi.org/10.1051/0004-6361/201321591}{{\em Astron. Astrophys.}
  {\bfseries 571} (2014) A16},
\href{http://arxiv.org/abs/1303.5076}{{\ttfamily arXiv:1303.5076
  [astro-ph.CO]}}.
\mciteBstWouldAddEndPunctfalse
\mciteSetBstMidEndSepPunct{\mcitedefaultmidpunct}
{}{\mcitedefaultseppunct}\relax
\EndOfBibitem
\bibitem{Williams:2012pz}
A.~J. Williams, C.~Boehm, S.~M. West, and D.~A. Vasquez, ``{Regenerating WIMPs
  in the Light of Direct and Indirect Detection},''
  \href{http://dx.doi.org/10.1103/PhysRevD.86.055018}{{\em Phys. Rev.}
  {\bfseries D86} (2012) 055018},
\href{http://arxiv.org/abs/1204.3727}{{\ttfamily arXiv:1204.3727 [hep-ph]}}.
\mciteBstWouldAddEndPunctfalse
\mciteSetBstMidEndSepPunct{\mcitedefaultmidpunct}
{}{\mcitedefaultseppunct}\relax
\EndOfBibitem
\bibitem{Navarro:1995iw}
J.~F. Navarro, C.~S. Frenk, and S.~D.~M. White, ``{The Structure of cold dark
  matter halos},'' \href{http://dx.doi.org/10.1086/177173}{{\em Astrophys. J.}
  {\bfseries 462} (1996) 563--575},
\href{http://arxiv.org/abs/astro-ph/9508025}{{\ttfamily arXiv:astro-ph/9508025
  [astro-ph]}}.
\mciteBstWouldAddEndPunctfalse
\mciteSetBstMidEndSepPunct{\mcitedefaultmidpunct}
{}{\mcitedefaultseppunct}\relax
\EndOfBibitem
\bibitem{Drees:2013wra}
M.~Drees, H.~Dreiner, D.~Schmeier, J.~Tattersall, and J.~S. Kim, ``{CheckMATE:
  Confronting your Favourite New Physics Model with LHC Data},''
  \href{http://dx.doi.org/10.1016/j.cpc.2014.10.018}{{\em Comput.Phys.Commun.}
  {\bfseries 187} (2014) 227--265},
\href{http://arxiv.org/abs/1312.2591}{{\ttfamily arXiv:1312.2591 [hep-ph]}}.
\mciteBstWouldAddEndPunctfalse
\mciteSetBstMidEndSepPunct{\mcitedefaultmidpunct}
{}{\mcitedefaultseppunct}\relax
\EndOfBibitem
\bibitem{Barr:2003rg}
A.~Barr, C.~Lester, and P.~Stephens, ``{m(T2): The Truth behind the glamour},''
  \href{http://dx.doi.org/10.1088/0954-3899/29/10/304}{{\em J.Phys.} {\bfseries
  G29} (2003) 2343--2363},
\href{http://arxiv.org/abs/hep-ph/0304226}{{\ttfamily arXiv:hep-ph/0304226
  [hep-ph]}}.
\mciteBstWouldAddEndPunctfalse
\mciteSetBstMidEndSepPunct{\mcitedefaultmidpunct}
{}{\mcitedefaultseppunct}\relax
\EndOfBibitem
\bibitem{Cheng:2008hk}
H.-C. Cheng and Z.~Han, ``{Minimal Kinematic Constraints and m(T2)},''
  \href{http://dx.doi.org/10.1088/1126-6708/2008/12/063}{{\em JHEP} {\bfseries
  0812} (2008) 063},
\href{http://arxiv.org/abs/0810.5178}{{\ttfamily arXiv:0810.5178 [hep-ph]}}.
\mciteBstWouldAddEndPunctfalse
\mciteSetBstMidEndSepPunct{\mcitedefaultmidpunct}
{}{\mcitedefaultseppunct}\relax
\EndOfBibitem
\bibitem{Cacciari:2005hq}
M.~Cacciari and G.~P. Salam, ``{Dispelling the $N^{3}$ myth for the $k_t$
  jet-finder},'' \href{http://dx.doi.org/10.1016/j.physletb.2006.08.037}{{\em
  Phys.Lett.} {\bfseries B641} (2006) 57--61},
\href{http://arxiv.org/abs/hep-ph/0512210}{{\ttfamily arXiv:hep-ph/0512210
  [hep-ph]}}.
\mciteBstWouldAddEndPunctfalse
\mciteSetBstMidEndSepPunct{\mcitedefaultmidpunct}
{}{\mcitedefaultseppunct}\relax
\EndOfBibitem
\bibitem{Cacciari:2008gp}
M.~Cacciari, G.~P. Salam, and G.~Soyez, ``{The Anti-k(t) jet clustering
  algorithm},'' \href{http://dx.doi.org/10.1088/1126-6708/2008/04/063}{{\em
  JHEP} {\bfseries 0804} (2008) 063},
\href{http://arxiv.org/abs/0802.1189}{{\ttfamily arXiv:0802.1189 [hep-ph]}}.
\mciteBstWouldAddEndPunctfalse
\mciteSetBstMidEndSepPunct{\mcitedefaultmidpunct}
{}{\mcitedefaultseppunct}\relax
\EndOfBibitem
\bibitem{Cacciari:2011ma}
M.~Cacciari, G.~P. Salam, and G.~Soyez, ``{FastJet User Manual},''
  \href{http://dx.doi.org/10.1140/epjc/s10052-012-1896-2}{{\em Eur.Phys.J.}
  {\bfseries C72} (2012) 1896},
\href{http://arxiv.org/abs/1111.6097}{{\ttfamily arXiv:1111.6097 [hep-ph]}}.
\mciteBstWouldAddEndPunctfalse
\mciteSetBstMidEndSepPunct{\mcitedefaultmidpunct}
{}{\mcitedefaultseppunct}\relax
\EndOfBibitem
\bibitem{deFavereau:2013fsa}
{\bfseries DELPHES 3} Collaboration, J.~de~Favereau {\em et~al.}, ``{DELPHES 3,
  A modular framework for fast simulation of a generic collider experiment},''
  \href{http://dx.doi.org/10.1007/JHEP02(2014)057}{{\em JHEP} {\bfseries 1402}
  (2014) 057},
\href{http://arxiv.org/abs/1307.6346}{{\ttfamily arXiv:1307.6346 [hep-ex]}}.
\mciteBstWouldAddEndPunctfalse
\mciteSetBstMidEndSepPunct{\mcitedefaultmidpunct}
{}{\mcitedefaultseppunct}\relax
\EndOfBibitem
\bibitem{Lester:1999tx}
C.~Lester and D.~Summers, ``{Measuring masses of semiinvisibly decaying
  particles pair produced at hadron colliders},''
  \href{http://dx.doi.org/10.1016/S0370-2693(99)00945-4}{{\em Phys.Lett.}
  {\bfseries B463} (1999) 99--103},
\href{http://arxiv.org/abs/hep-ph/9906349}{{\ttfamily arXiv:hep-ph/9906349
  [hep-ph]}}.
\mciteBstWouldAddEndPunctfalse
\mciteSetBstMidEndSepPunct{\mcitedefaultmidpunct}
{}{\mcitedefaultseppunct}\relax
\EndOfBibitem
\bibitem{Read:2002hq}
A.~L. Read, ``{Presentation of search results: The CL(s) technique},''
\href{http://dx.doi.org/10.1088/0954-3899/28/10/313}{{\em J.Phys.} {\bfseries
  G28} (2002) 2693--2704}.
\mciteBstWouldAddEndPunctfalse
\mciteSetBstMidEndSepPunct{\mcitedefaultmidpunct}
{}{\mcitedefaultseppunct}\relax
\EndOfBibitem
\bibitem{Agashe:2014kda}
{\bfseries Particle Data Group} Collaboration, K.~A. Olive {\em et~al.},
  ``{Review of Particle Physics},''
\href{http://dx.doi.org/10.1088/1674-1137/38/9/090001}{{\em Chin. Phys.}
  {\bfseries C38} (2014) 090001}.
\mciteBstWouldAddEndPunctfalse
\mciteSetBstMidEndSepPunct{\mcitedefaultmidpunct}
{}{\mcitedefaultseppunct}\relax
\EndOfBibitem
\bibitem{bsgamma}
{\url{http://www.slac.stanford.edu/xorg/hfag/rare/2012/radll/index.html}}\relax
\mciteBstWouldAddEndPunctfalse
\mciteSetBstMidEndSepPunct{\mcitedefaultmidpunct}
{}{\mcitedefaultseppunct}\relax
\EndOfBibitem
\bibitem{Adachi:2012mm}
{\bfseries Belle} Collaboration, I.~Adachi {\em et~al.}, ``{Evidence for $B^-
  \to \tau^- \bar{\nu}_\tau$ with a Hadronic Tagging Method Using the Full Data
  Sample of Belle},''
  \href{http://dx.doi.org/10.1103/PhysRevLett.110.131801}{{\em Phys.Rev.Lett.}
  {\bfseries 110} no.~13, (2013) 131801},
\href{http://arxiv.org/abs/1208.4678}{{\ttfamily arXiv:1208.4678 [hep-ex]}}.
\mciteBstWouldAddEndPunctfalse
\mciteSetBstMidEndSepPunct{\mcitedefaultmidpunct}
{}{\mcitedefaultseppunct}\relax
\EndOfBibitem
\bibitem{Aaij:2013aka}
{\bfseries LHCb} Collaboration, R.~Aaij {\em et~al.}, ``{Measurement of the
  $B^0_s \to \mu^+ \mu^-$ branching fraction and search for $B^0 \to \mu^+
  \mu^-$ decays at the LHCb experiment},''
  \href{http://dx.doi.org/10.1103/PhysRevLett.111.101805}{{\em Phys.Rev.Lett.}
  {\bfseries 111} (2013) 101805},
\href{http://arxiv.org/abs/1307.5024}{{\ttfamily arXiv:1307.5024 [hep-ex]}}.
\mciteBstWouldAddEndPunctfalse
\mciteSetBstMidEndSepPunct{\mcitedefaultmidpunct}
{}{\mcitedefaultseppunct}\relax
\EndOfBibitem
\bibitem{Chatrchyan:2013bka}
{\bfseries CMS} Collaboration, S.~Chatrchyan {\em et~al.}, ``{Measurement of
  the B(s) to mu+ mu- branching fraction and search for B0 to mu+ mu- with the
  CMS Experiment},''
  \href{http://dx.doi.org/10.1103/PhysRevLett.111.101804}{{\em Phys.Rev.Lett.}
  {\bfseries 111} (2013) 101804},
\href{http://arxiv.org/abs/1307.5025}{{\ttfamily arXiv:1307.5025 [hep-ex]}}.
\mciteBstWouldAddEndPunctfalse
\mciteSetBstMidEndSepPunct{\mcitedefaultmidpunct}
{}{\mcitedefaultseppunct}\relax
\EndOfBibitem
\bibitem{Kowalska:2012gs}
K.~Kowalska {\em et~al.}, ``{Constrained next-to-minimal supersymmetric
  standard model with a 126 GeV Higgs boson: A global analysis},''
  \href{http://dx.doi.org/10.1103/PhysRevD.87.115010}{{\em Phys.Rev.}
  {\bfseries D87} no.~11, (2013) 115010},
\href{http://arxiv.org/abs/1211.1693}{{\ttfamily arXiv:1211.1693 [hep-ph]}}.
\mciteBstWouldAddEndPunctfalse
\mciteSetBstMidEndSepPunct{\mcitedefaultmidpunct}
{}{\mcitedefaultseppunct}\relax
\EndOfBibitem
\bibitem{Fowlie:2013oua}
A.~Fowlie, K.~Kowalska, L.~Roszkowski, E.~M. Sessolo, and Y.-L.~S. Tsai,
  ``{Dark matter and collider signatures of the MSSM},''
  \href{http://dx.doi.org/10.1103/PhysRevD.88.055012}{{\em Phys.Rev.}
  {\bfseries D88} (2013) 055012},
\href{http://arxiv.org/abs/1306.1567}{{\ttfamily arXiv:1306.1567 [hep-ph]}}.
\mciteBstWouldAddEndPunctfalse
\mciteSetBstMidEndSepPunct{\mcitedefaultmidpunct}
{}{\mcitedefaultseppunct}\relax
\EndOfBibitem
\bibitem{Feroz:2008xx}
F.~Feroz, M.~Hobson, and M.~Bridges, ``{MultiNest: an efficient and robust
  Bayesian inference tool for cosmology and particle physics},''
  \href{http://dx.doi.org/10.1111/j.1365-2966.2009.14548.x}{{\em
  Mon.Not.Roy.Astron.Soc.} {\bfseries 398} (2009) 1601--1614},
\href{http://arxiv.org/abs/0809.3437}{{\ttfamily arXiv:0809.3437 [astro-ph]}}.
\mciteBstWouldAddEndPunctfalse
\mciteSetBstMidEndSepPunct{\mcitedefaultmidpunct}
{}{\mcitedefaultseppunct}\relax
\EndOfBibitem
\bibitem{Allanach:2001kg}
B.~Allanach, ``{SOFTSUSY: a program for calculating supersymmetric spectra},''
  \href{http://dx.doi.org/10.1016/S0010-4655(01)00460-X}{{\em
  Comput.Phys.Commun.} {\bfseries 143} (2002) 305--331},
\href{http://arxiv.org/abs/hep-ph/0104145}{{\ttfamily arXiv:hep-ph/0104145
  [hep-ph]}}.
\mciteBstWouldAddEndPunctfalse
\mciteSetBstMidEndSepPunct{\mcitedefaultmidpunct}
{}{\mcitedefaultseppunct}\relax
\EndOfBibitem
\bibitem{Mahmoudi:2008tp}
F.~Mahmoudi, ``{SuperIso v2.3: A Program for calculating flavor physics
  observables in Supersymmetry},''
  \href{http://dx.doi.org/10.1016/j.cpc.2009.02.017}{{\em Comput.Phys.Commun.}
  {\bfseries 180} (2009) 1579--1613},
\href{http://arxiv.org/abs/0808.3144}{{\ttfamily arXiv:0808.3144 [hep-ph]}}.
\mciteBstWouldAddEndPunctfalse
\mciteSetBstMidEndSepPunct{\mcitedefaultmidpunct}
{}{\mcitedefaultseppunct}\relax
\EndOfBibitem
\bibitem{Belanger:2013oya}
G.~Belanger, F.~Boudjema, A.~Pukhov, and A.~Semenov, ``{micrOMEGAs 3: A program
  for calculating dark matter observables},''
  \href{http://dx.doi.org/10.1016/j.cpc.2013.10.016}{{\em Comput.Phys.Commun.}
  {\bfseries 185} (2014) 960--985},
\href{http://arxiv.org/abs/1305.0237}{{\ttfamily arXiv:1305.0237 [hep-ph]}}.
\mciteBstWouldAddEndPunctfalse
\mciteSetBstMidEndSepPunct{\mcitedefaultmidpunct}
{}{\mcitedefaultseppunct}\relax
\EndOfBibitem
\bibitem{ATLAS:2014wva}
{\bfseries ATLAS, CDF, CMS, D0} Collaboration, ``{First combination of Tevatron
  and LHC measurements of the top-quark mass},''
\href{http://arxiv.org/abs/1403.4427}{{\ttfamily arXiv:1403.4427 [hep-ex]}}.
\mciteBstWouldAddEndPunctfalse
\mciteSetBstMidEndSepPunct{\mcitedefaultmidpunct}
{}{\mcitedefaultseppunct}\relax
\EndOfBibitem
\bibitem{Catena:2009mf}
R.~Catena and P.~Ullio, ``{A novel determination of the local dark matter
  density},'' \href{http://dx.doi.org/10.1088/1475-7516/2010/08/004}{{\em JCAP}
  {\bfseries 1008} (2010) 004},
\href{http://arxiv.org/abs/0907.0018}{{\ttfamily arXiv:0907.0018
  [astro-ph.CO]}}.
\mciteBstWouldAddEndPunctfalse
\mciteSetBstMidEndSepPunct{\mcitedefaultmidpunct}
{}{\mcitedefaultseppunct}\relax
\EndOfBibitem
\bibitem{Bovy:2012tw}
J.~Bovy and S.~Tremaine, ``{On the local dark matter density},''
  \href{http://dx.doi.org/10.1088/0004-637X/756/1/89}{{\em Astrophys. J.}
  {\bfseries 756} (2012) 89},
\href{http://arxiv.org/abs/1205.4033}{{\ttfamily arXiv:1205.4033
  [astro-ph.GA]}}.
\mciteBstWouldAddEndPunctfalse
\mciteSetBstMidEndSepPunct{\mcitedefaultmidpunct}
{}{\mcitedefaultseppunct}\relax
\EndOfBibitem
\bibitem{Read:2014qva}
J.~I. Read, ``{The Local Dark Matter Density},''
  \href{http://dx.doi.org/10.1088/0954-3899/41/6/063101}{{\em J. Phys.}
  {\bfseries G41} (2014) 063101},
\href{http://arxiv.org/abs/1404.1938}{{\ttfamily arXiv:1404.1938
  [astro-ph.GA]}}.
\mciteBstWouldAddEndPunctfalse
\mciteSetBstMidEndSepPunct{\mcitedefaultmidpunct}
{}{\mcitedefaultseppunct}\relax
\EndOfBibitem
\bibitem{Savage:2015xta}
C.~Savage, A.~Scaffidi, M.~White, and A.~G. Williams, ``{LUX likelihood and
  limits on spin-independent and spin-dependent WIMP couplings with LUXCalc},''
\href{http://arxiv.org/abs/1502.02667}{{\ttfamily arXiv:1502.02667 [hep-ph]}}.
\mciteBstWouldAddEndPunctfalse
\mciteSetBstMidEndSepPunct{\mcitedefaultmidpunct}
{}{\mcitedefaultseppunct}\relax
\EndOfBibitem
\bibitem{2012arXiv1206.6288A}
E.~{Aprile} and {XENON1T collaboration}, ``{The XENON1T Dark Matter Search
  Experiment},'' {\em ArXiv e-prints} (June, 2012) ,
  \href{http://arxiv.org/abs/1206.6288}{{\ttfamily arXiv:1206.6288
  [astro-ph.IM]}}\relax
\mciteBstWouldAddEndPunctfalse
\mciteSetBstMidEndSepPunct{\mcitedefaultmidpunct}
{}{\mcitedefaultseppunct}\relax
\EndOfBibitem
\bibitem{Cirelli:2010xx}
M.~Cirelli, G.~Corcella, A.~Hektor, G.~Hutsi, M.~Kadastik, P.~Panci, M.~Raidal,
  F.~Sala, and A.~Strumia, ``{PPPC 4 DM ID: A Poor Particle Physicist Cookbook
  for Dark Matter Indirect Detection},''
  \href{http://dx.doi.org/10.1088/1475-7516/2012/10/E01,
  10.1088/1475-7516/2011/03/051}{{\em JCAP} {\bfseries 1103} (2011) 051},
  \href{http://arxiv.org/abs/1012.4515}{{\ttfamily arXiv:1012.4515 [hep-ph]}}.
[Erratum: JCAP1210,E01(2012)].
\mciteBstWouldAddEndPunctfalse
\mciteSetBstMidEndSepPunct{\mcitedefaultmidpunct}
{}{\mcitedefaultseppunct}\relax
\EndOfBibitem
\bibitem{Akula:2013ioa}
S.~Akula and P.~Nath, ``{Gluino-driven radiative breaking, Higgs boson mass,
  muon g-2, and the Higgs diphoton decay in supergravity unification},''
  \href{http://dx.doi.org/10.1103/PhysRevD.87.115022}{{\em Phys. Rev.}
  {\bfseries D87} no.~11, (2013) 115022},
\href{http://arxiv.org/abs/1304.5526}{{\ttfamily arXiv:1304.5526 [hep-ph]}}.
\mciteBstWouldAddEndPunctfalse
\mciteSetBstMidEndSepPunct{\mcitedefaultmidpunct}
{}{\mcitedefaultseppunct}\relax
\EndOfBibitem
\bibitem{Nath:2015dza}
P.~Nath, ``{Supersymmetry after the Higgs},''
  \href{http://dx.doi.org/10.1002/andp.201500005}{{\em Annalen Phys.} (2015)
  1--12},
\href{http://arxiv.org/abs/1501.01679}{{\ttfamily arXiv:1501.01679 [hep-ph]}}.
\mciteBstWouldAddEndPunctfalse
\mciteSetBstMidEndSepPunct{\mcitedefaultmidpunct}
{}{\mcitedefaultseppunct}\relax
\EndOfBibitem
\bibitem{Kowalska:2015zja}
K.~Kowalska, L.~Roszkowski, E.~M. Sessolo, and A.~J. Williams, ``{GUT-inspired
  SUSY and the muon g − 2 anomaly: prospects for LHC 14 TeV},''
  \href{http://dx.doi.org/10.1007/JHEP06(2015)020}{{\em JHEP} {\bfseries 06}
  (2015) 020},
\href{http://arxiv.org/abs/1503.08219}{{\ttfamily arXiv:1503.08219 [hep-ph]}}.
\mciteBstWouldAddEndPunctfalse
\mciteSetBstMidEndSepPunct{\mcitedefaultmidpunct}
{}{\mcitedefaultseppunct}\relax
\EndOfBibitem
\end{mcitethebibliography}
